\documentclass[twocolumn]{aastex7}

\usepackage{verbatim}
\usepackage{amsmath}
\usepackage{subfigure}
\usepackage{hyperref}
\usepackage{graphicx}
\usepackage{color}

\def\E{{\cal E}}
\def\F{{\cal F}}
\def\R{{\cal R}}
\def\M{{\cal M}}
\def\bhat{\hat{\textbf{b}}}

\def\identity{{\sf I}}
\def\ptensor{{\sf P}}
\def\vtensor{{\sf \Pi}}

\newcommand{\code}[1]{\texttt{#1}}

\begin{document}

\title{Reduced Effective Viscosity from Anisotropic Transport and Plasma Instabilities in the Sloshing Cores of Galaxy Clusters}

\author[orcid=0000-0003-3175-2347]{John A. ZuHone}
\affiliation{Center for Astrophysics $\vert$ Harvard \& Smithsonian, 60 Garden Street, Cambridge, MA 02138, USA}
\email[show]{john.zuhone@cfa.harvard.edu}  

\author[orcid=0000-0002-7879-060X]{Stephen Majeski}
\affiliation{JILA, University of Colorado and National Institute of Standards and Technology, 440 UCB, Boulder, CO 80309, USA}
\affiliation{Department of Astrophysical Sciences, Princeton University, Peyton Hall, Princeton, NJ 08544, USA}
\email{stephen.majeski@colorado.edu}

\author[orcid=0000-0002-7726-4202]{Annie Heinrich}
\affiliation{Department of Astronomy and Astrophysics, The University of Chicago, Chicago, IL 60637, USA}
\email{amheinrich@uchicago.edu}

\author[orcid=0000-0003-4421-1128]{Alexander A. Schekochihin}
\affiliation{Rudolf Peierls Centre for Theoretical Physics, University of Oxford, Oxford OX1 3PU, UK}
\affiliation{Merton College, Oxford OX1 4JD, UK}
\email{a.schekochihin1@physics.ox.ac.uk}

\author[orcid=0000-0002-8820-8177]{Francisco Ley}
\affiliation{Department of Astronomy, University of Wisconsin-Madison, Madison, WI 53706, USA}
\email{fley.astro@gmail.com}

\author[orcid=0000-0001-7630-8085]{Irina Zhuravleva}
\affiliation{Department of Astronomy and Astrophysics, The University of Chicago, Chicago, IL 60637, USA}
\email{zhuravleva@uchicago.edu}

\begin{abstract}
The $\sim \mu$G magnetic field in the intracluster medium (ICM) introduces a pressure anisotropy with respect to the
magnetic field's direction that manifests as an anisotropic viscous stress. Plasma instabilities arising from the
pressure anisotropy crossing certain thresholds force it to marginally stable values, reducing viscous transport.
Additionally, the feedback of this anisotropic pressure on the velocity field has been predicted to lead to a form of
self-organization that also can reduce viscous dissipation without affecting the collisionality. In this work, we
present high-resolution Braginskii-MHD simulations of a galaxy cluster core with sloshing gas motions and turbulence,
including the effects of anisotropic viscous stress and different simple prescriptions for limiting the pressure
anisotropy due to plasma instabilities. Braginskii viscosity has an expected, though modest, effect on suppressing
Kelvin-Helmholtz instabilities at sloshing cold front surfaces, dependent on how the pressure anisotropy is limited.
Due to the sloshing motions, the magnetic field's strength can become high enough in places that the pressure anisotropy 
need not be limited. Nevertheless, the combined effect of the limiters and the turbulent structure of the magnetic field in
all simulations is that the effective viscosity is much lower than the isotropic Spitzer value in a significant fraction
of the core region. However, we find that this reduced viscosity is capable of steepening the velocity-amplitude
spectrum and transferring a small fraction of the turbulent kinetic energy into heat. Finally, we present evidence 
for magneto-immutable dynamics in our simulations.
\end{abstract}


\section{Introduction}\label{sec:intro}

The largest example of a hot space plasma is the intracluster medium (ICM), the X-ray-emitting diffuse gas that fills
the dark matter (DM) dominated gravitational well of galaxy clusters. Space-based observatories such as {\it Chandra},
{\it XMM-Newton}, {\it Suzaku}, {\it NuSTAR}, and {\it eROSITA} have provided detailed X-ray images of the ICM at
moderate spectral resolution, which have revealed a wealth of structure and complexity in the gas distribution in terms
of its density, temperature, and metallicity. More recently, the high-spectral resolution microcalorimeter instruments
aboard {\it Hitomi} and {\it XRISM} have provided the first detailed spatially resolved measurements of the velocity
structure of the ICM in a number of clusters. It is also well-established from radio observations that the ICM is weakly magnetized, with observed field strengths of $\sim$1-10~$\mu$G \citep{Carilli2002,Ferrari2008,Bruggen2013,Vacca2018}. This corresponds to $\beta = p_{\rm th}/p_B \sim 100$, so the magnetic pressure associated with this field is much smaller than the thermal pressure (although it is comparable to the energy density of the turbulent motions).

It is frequently assumed in the literature that the ICM plasma can be reasonably described as a fluid, because the mean
free path $\lambda_{\rm mfp}$ of its particles is shorter than the typical length scales of the system $L$, although not
extremely so. To quantify this, we assume that $\lambda_{\rm mfp}$ arises from Coulomb collisions between the electrons
and ions (either between themselves or each other) and can hence estimate it as \citep{Spitzer1962}:
\begin{equation}\label{eqn:cmfp}
\lambda_{\rm mfp} = \frac{3^{3/2}(k_BT)^2}{4\pi^{1/2}ne^4\ln{\Lambda}},
\end{equation}
where $n$ is the electron/ion number density, $T$ is the gas temperature, $e$ is the elementary unit of charge, and
$\ln\Lambda$ is the ``Coulomb logarithm'', where $\Lambda$ is the ratio of the largest to the smallest impact parameters
for particle collisions, and has typical values of $\sim$30-40. In the cores of relaxed, cool-core galaxy clusters,
$\lambda_{\rm mfp}$ can be less than 1~kpc, and can rise to 10s of kpc at larger radii. As for $L$, it depends on which
length scale is under consideration. If it is the cluster core, then $L \sim 30-100$~kpc, but alternatively we may
consider the size of the cluster as a whole, in which case $L \sim 1-2$~Mpc. 

Given that $\lambda_{\rm mfp} \ll L$, most simulations that have sought to capture the ICM properties and explain
its observed features have been performed in the fluid approximation. Nevertheless,
questions remain about the microphysical properties of the ICM that depend more sensitively on the precise details of
its collisionality, such as the diffusive transport of momentum (viscosity) and heat (conductivity). Furthermore, while the condition $\lambda_{\rm mfp} \ll L$ allows one to derive a fluid theory via the Chapman-Enskog expansion, it is not \textit{a priori} guaranteed---and in a high-$\beta$ plasma such as the ICM, generally not true---that the fluid solutions are stable to microscale, collisionless perturbations \citep{Bott2024}.

Assuming that Coulomb collisions are the dominant source of momentum transport in the ICM, the dynamical viscosity is
given by the familiar ``Spitzer'' \citep{Spitzer1962,Braginskii1965,Sarazin1988} formula:
\begin{eqnarray}
\nonumber \mu &\approx& \frac{1}{3}n_im_i{v_{{\rm rms},i}}\lambda_i \label{eqn:spitzer_visc} \\
\nonumber &=& 0.58n_im_i\left(\frac{k_BT}{m_i}\right)^{1/2}\lambda_i \\
 &\approx& 1410~{\rm g}~{\rm s}^{-1}~{\rm cm}^{-1}\left(\frac{k_BT}{5~{\rm
keV}}\right)^{5/2}\left(\frac{\ln{\Lambda}}{40}\right)^{-1},
\end{eqnarray}
where $m_i$ is the mass of the ions, $n_i$ is the ion number density, and $\lambda_i$ is the mean free path of the ions.
The viscous transport of momentum by the electrons can be neglected because $m_e \ll m_i$. Assuming conditions
appropriate for a cluster core and the definition of the kinematic viscosity as $\nu = \mu/\rho$, under
the above assumptions we can estimate the Reynolds number of the ICM as follows:
\begin{eqnarray}
{\rm Re} &=& \frac{vL}{\nu} \label{eqn:reynolds} \\
\nonumber &\sim& 100\left(\frac{v}{300~{\rm km~s}^{-1}}\right)\left(\frac{L}{30~{\rm
kpc}}\right)\left(\frac{k_BT}{4~{\rm keV}}\right)^{-5/2} \\
\nonumber &\times& \left(\frac{\ln{\Lambda}}{40}\right)\left(\frac{n_i}{10^{-2}~\rm{cm}^{-3}}\right),
\end{eqnarray}
where $v$ is the gas velocity. Given that the density, temperature, and gas velocity can vary across cluster cores, this
value is uncertain, ${\rm Re} \sim 1-200$. Nonetheless, this rough value indicates that the effects
of viscosity could be relevant to ICM macrophysics if the momentum transport is mediated by Coulomb collisions in an
isotropic fashion. 

However, there is observational evidence as well as theoretical considerations to suggest that the viscosity of the ICM
is lower than that predicted by Equations \ref{eqn:spitzer_visc} and \ref{eqn:reynolds}. Many clusters host ``cold
fronts'' (hereafter CFs), which are subsonically moving contact discontinuities (pressure-continuous, where the denser
side of the interface is colder) that appear where there is ongoing or recent merger activity. These
manifest most often either in ``major'' cluster mergers between two systems of comparable mass \citep[such as the
``Bullet Cluster'', see][]{Markevitch2002}, or ``sloshing'' motions in the cores of massive galaxy clusters produced by
interactions with small subclusters \citep[such as those observed in the Perseus Cluster, but also in many other
systems; see][]{Markevitch2000,Markevitch2001,Churazov2003,Simionescu2007,Simionescu2010,Fabian2011,Rossetti2013,Walker2018}. Hydrodynamic simulations of the formation of CFs in merging clusters have shown that these
features are also the sites of velocity shear flows, which should readily produce Kelvin-Helmholtz instabilities
\citep[hereafter KHI;][]{ZuHone2010,Roediger2011}. These either appear as wave-like features along the front surfaces or
as ripples in the X-ray surface brightness (hereafter SB) profiles behind the fronts in projection \citep{Roediger2013}.
The first observations of CFs with {\it Chandra} indicated that they appeared relatively smooth over large length
scales, with no strong evidence of KHI, indicating that the growth of the instabilities may be suppressed by some
mechanism. Later on in the mission, longer exposures of the same and other CFs providing better statistics demonstrated
conclusively the presence of features which were consistent with KHI \citep{Werner2016,Su2017,Wang2018}. These studies
estimated that the effective ICM viscosity is at most $\sim$5-20\% of the Spitzer value. 

\citet{Roediger2015a,Roediger2015b} provided evidence that the relatively short length of ram-pressure stripped tails
behind elliptical galaxies falling into clusters (such as M89) indicates a low viscosity (at most $\sim$10\% of
Spitzer), as the tail gas is instead mixed by turbulence into the surrounding ICM. Corroborating and independent
evidence for such low-viscosity, stripped turbulent tails was provided by \citet{Li2023} and \citet{Ignesti2024}.
Analysis of \textit{Chandra} observations of SB fluctuations down to small length scales in the Coma
cluster by \citet{Zhuravleva2019} and \citet{Heinrich2024} provided indirect measurements of the velocity power spectrum
in the ICM; comparison to simulations indicated that the inertial range of turbulence in the ICM extends to small length
scales and that the implied viscosity is at most $\sim$1-5\% of the Spitzer value.

In the past 10 years, the advent of spatially resolved high-resolution X-ray spectroscopy with microcalorimeter
instruments has enabled the first direct measurements of the velocity structure of the ICM via Doppler broadening and
shifting of emission lines. The {\it Hitomi} spacecraft made the first such measurements in the Perseus Cluster
\citep{Hitomi2016,Hitomi2018}. Despite evidence for a dynamic ICM from AGN cavities and sloshing CFs, the bulk motions
and velocity dispersions are on the order of $\lesssim 200$~km~s$^{-1}$, which implies non-thermal pressure support of
$\lesssim 10$\% of the thermal pressure in the core region. After its recent launch, the {\it XRISM} spacecraft has
continued these measurements in a number of clusters, including the Perseus cluster \citep{XRISMPerseus2026}, as well as several other systems \citep[e.g.][]{XRISMComa,XRISMCentaurus,XRISMA2029a,XRISMA2029b}, with a very
similar result for nearly all of the observed clusters so far---the velocity dispersions implied by velocity broadening
are small compared to thermal broadening, i.e., the ICM turbulence is subsonic. It has sometimes been claimed in the
literature that such small velocities may be due to significant viscosity, though some works using mock {\it Hitomi} and
{\it XRISM} observations of simulations of the ICM have indicated that a large viscosity is not necessary to explain
such small velocity dispersions \citep[e.g.][]{ZuHone2016,Bourne2017,Lau2017,ZuHone2018,Ehlert2021,Truong2024}.

There are also theoretical considerations that suggest the viscosity of the ICM should be low. Magnetic fields have a
significant effect on the ICM transport processes due to the fact that the Larmor radii of the ions and electrons are
much smaller than their Coulomb mean free paths. The Larmor radius of the ions is given by:
\begin{equation}\label{eqn:larmor}
\rho_L = \frac{m_icv_\perp}{eB} \approx 3 \times 10^{-12}~{\rm kpc} \left(\frac{k_BT}{5~{\rm keV}}\right)^{1/2}\left(\frac{B}{1~{\rm \mu{G}}}\right)^{-1},
\end{equation}
where $m_i$ is the mass of the ion, $c$ is the speed of light, $v_\perp$ is the component of the particle velocity
perpendicular to the magnetic field, and $B$ is the magnetic-field strength. The most probable value for $v_\perp$,
assuming a Maxwellian velocity distribution, is $v_\perp = \sqrt{2k_BT/m_i}$. The large difference in scales between
$\rho_L$ and $\lambda_{\rm mfp}$ leads to the transport of momentum being strongly anisotropic, which is known as
Braginskii viscosity \citep{Braginskii1965}. Depending on the local direction of the magnetic field with respect to
velocity gradients, this may result in a significant reduction in the effective viscosity compared to the isotropic
case. A number of previous studies have included the effects of Braginskii viscosity in MHD simulations of
various situations in the ICM, including rising AGN-driven bubbles \citep{Dong2009,Kingsland2019}, merger and
sloshing-driven CFs \citep{Suzuki2013,ZuHone2015}, and high-resolution treatments of the development of fluid
instabilities such as KHI \citep{Berlok2019,Berlok2020}. In each of these cases, the effect of the Braginskii viscosity
on quantities that can be probed by X-ray observations is to reduce the prevalence of KHI to a
degree similar to that produced by an isotropic viscosity that is $\sim$5-20\% of the Spitzer value, depending on the field-line
geometry and conditions of the plasma. 

The anisotropy of the particle motion with respect to the magnetic field's direction also results in an anisotropy in the
pressure of electrons and ions \citep{Chew1956}. For charged particles gyrating around magnetic-field lines that are
changing at a lower rate than the gyrofrequency $\Omega = \rho_L/v_\perp$, the magnetic moment $\mu_B = m_iv_\perp/2B$
is an adiabatic invariant. Thus, slow changes in $B$ result in adiabatic changes in the perpendicular velocity $v_\perp$
of the particles. A separate adiabatic invariant applies to the parallel velocity $v_\parallel$, associated with
``bouncing'' of particles in local ``magnetic-bottle'' fluctuations. Since in general the particle speeds in the
perpendicular and parallel directions will not change in the same way, it is to be expected that there will be some
degree of pressure anisotropy, defined as the difference between the perpendicular ($p_\perp$) and parallel
($p_\parallel$) pressures as $\Delta{p} = p_\perp - p_\parallel$, with respect to the local magnetic-field direction in
the ICM. While the changing magnetic field will increase the pressure anisotropy, Coulomb collisions will act to reduce
it, so the value of the anisotropy at any given location will depend on the local characteristics of the magnetic field
and the kinematics and thermodynamics of the plasma. As we will show in detail in Section \ref{sec:physics}, the Braginskii stress itself can be expressed in terms of the pressure anisotropy. 

It is at this point that other interesting effects come into play, which are related to the fact that in the
high-$\beta$ ICM, the viscous stress and the stress from the magnetic tension can easily become comparable in magnitude.
Roughly speaking, as the absolute value of the pressure anisotropy $\Delta{p}$ exceeds the magnetic pressure $B^2/4\pi$,
the plasma can become unstable to rapidly growing kinetic instabilities, which drive fluctuations in the magnetic field
at the Larmor scale. The ions then pitch-angle scatter off these fluctuations, which partially isotropizes their
velocities and reduces the pressure anisotropy to marginally stable values. The upshot is that the effective
collisionality of the ICM becomes less dependent on the particle-particle Coulomb interactions described by Equation
\ref{eqn:cmfp} and is instead determined by wave-particle interactions, the result of which is an effective
suppression of viscous stress. We will discuss these issues in more detail in Section \ref{sec:physics}; a thorough
recent review is given by \citet[][see their Sections 2.1-2.5]{Kunz2022}.

In this work, motivated by these considerations, we explore the effects on the ICM of Braginskii viscosity, adjusted by
a simple prescription for limiting the pressure anisotropy to account for the effects of microscale plasma
instabilities. The scenario that we adopt is that of a cool-core cluster with sloshing gas motions, which applies to
many observed systems (including clusters such as Perseus, Centaurus, and Abell~2029, all of which have not only been
extensively observed by {\it Chandra} and {\it XMM-Newton} but were among the first clusters to be observed by {\it
XRISM}). The sloshing motions produce CFs and drive a turbulent cascade, which makes a scenario such as this an ideal
choice to test the observable effects of viscosity in the ICM on the thermodynamics and kinematics of the gas. We expand
upon the work of \citet{ZuHone2015} by including the effects of ``hard-wall'' limiters on the pressure anisotropy
motivated by microscale plasma instabilities, as well as exploring different initial magnetic field strengths. 

This paper is organized as follows. In Section \ref{sec:method}, we describe the simulations and the code. In Section
\ref{sec:results}, we describe the effects of anisotropic viscosity. Finally, in Section \ref{sec:conclusions}, we
summarize our results and discuss future developments of this work. While this manuscript was being prepared, a similar study was performed by \citet{Hsieh2026}, which also explores the effects of viscosity and pressure-anisotropy limiters in simulations of sloshing CFs. We will make brief remarks on the comparison of our simulations to theirs in Section \ref{sec:conclusions}.

\section{Methods}\label{sec:method}

\subsection{Equations}\label{sec:physics}

Our simulations solve the following set of MHD equations, listed here in conservative form and assuming Gaussian
electromagnetic units:
\begin{eqnarray}
\frac{\partial{\rho}}{\partial{t}} + \nabla \cdot (\rho{\bf v}) &=& 0, \label{eqn:density}\\
\frac{\partial{(\rho{\bf v})}}{\partial{t}} + \nabla \cdot
\left(\rho{\bf vv} + p\identity - \frac{\bf BB}{4\pi} + \vtensor \right) &=&
\rho{\bf g} \label{eqn:momentum}, \\
\frac{\partial{E}}{\partial{t}} + \nabla \cdot \left\{\left[(E+p)\identity - \frac{{\bf
 B}{\bf B}}{4\pi} + \vtensor \right] \cdot {\bf v} \right\} &=& \rho{\bf g \cdot v},  \\
\frac{\partial{\bf B}}{\partial{t}} + \nabla \cdot ({\bf vB} - {\bf
 Bv}) &=& 0, \label{eqn:bfield}
\end{eqnarray}
where the total pressure $p$, thermal pressure $p_{\rm th}$, and total energy density $E$ are defined as $p = p_{\rm th} + p_B$, $p_{\rm th} = (\gamma - 1)\rho\epsilon$, and $E = \rho{v^2}/2 + \epsilon + p_B$. $\rho$ is the gas density, $\bf{v}$ is the fluid velocity, $\epsilon$ is the gas internal energy per unit mass, $\bf{B}$
is the magnetic field vector, $p_B = B^2/8\pi$ is the magnetic pressure, and $\identity$ is the unit dyadic. We assume
an ideal equation of state with $\gamma = 5/3$, equal electron and ion temperatures, and primordial abundances with
molecular weight $\bar{A} = 0.6$. 

Finally, $\vtensor$ is the viscous-stress tensor. In hydrodynamics, this tensor has the form
\begin{equation}
\vtensor = -\mu\left[\nabla{\bf v} + (\nabla{\bf v})^\mathsf{T} - \frac{2}{3}(\nabla\cdot{\bf v})\identity\right].
\label{eqn:visc_tensor_hydro}
\end{equation}
However, given that in the ICM $\lambda_{\rm mfp} \gg \rho_L$, the viscous flux is expected to be anisotropic with
respect to the local magnetic field's direction. Projecting velocity gradients into components perpendicular and parallel
to the magnetic field and noting that the components of the stress tensor that affect the momentum transport
parallel to the field lines are much larger than those perpendicular to them, we find that the viscous-stress
tensor instead takes the form \citep{Braginskii1965}
\begin{equation}
\vtensor = -3\mu\left(\bhat\bhat-\frac{1}{3}\identity\right)\left(\bhat\bhat-\frac{1}{3}\identity\right):\nabla{\bf v},
\label{eqn:visc_tensor_braginskii}
\end{equation}
where $\bhat = {\bf B}/B$ is the unit vector in the direction of the magnetic field. 

Let us explain how this anisotropic (Braginskii) viscosity in the ICM relates to the pressure anisotropy. As detailed in Section \ref{sec:intro}, the large scale separation between the Coulomb mean free paths and the Larmor
radii of the ions makes the pressure anisotropy $\Delta{p}$ physically relevant. The total thermal pressure satisfies
\begin{equation}
p_{\rm th} = \frac{2}{3} \,p_\perp+\frac{1}{3} \,p_\parallel.
\label{eqn:total_pressure}
\end{equation}
Differences between $p_\perp$ and $p_\parallel$ arise from the conservation of the first and second
adiabatic invariants for each particle on timescales much greater than the inverse of the ion gyrofrequency,
$\Omega_i^{-1}$ \citep{Chew1956}. Given these facts, the generic form of Equation \ref{eqn:momentum} is
\begin{equation}
\frac{\partial{(\rho{\bf v})}}{\partial{t}} + \nabla \cdot
\left(\rho{\bf vv} + \ptensor\right) = \rho{\bf g},
\label{eqn:momentum_alt}
\end{equation}
where the pressure tensor may be expressed as:
\begin{eqnarray}
\nonumber \ptensor &=& p_\parallel\bhat\bhat + p_\perp(\identity - \bhat\bhat) + p_B\identity - \frac{{\bf B}{\bf B}}{4\pi}\label{eqn:stress_tensor} \\
 &=& p\identity - \frac{{\bf B}{\bf B}}{4\pi} - \Delta{p}\left(\bhat\bhat-\frac{1}{3}\identity\right),
\end{eqnarray}
\noindent
where we have used Equation \ref{eqn:total_pressure}. By comparing Equation \ref{eqn:stress_tensor} with Equation \ref{eqn:momentum}, one immediately identifies the term proportional to the pressure anisotropy with the viscous stress tensor: 
\begin{equation}
\vtensor = -\Delta{p}\left(\bhat\bhat-\frac{1}{3}\identity\right).
\label{eqn:visc_stress}
\end{equation}

When the ion-ion collision frequency $\nu_{\rm ii}$ is much larger than the rates of
change of all fields, an equation for the pressure anisotropy can be obtained by balancing its production by adiabatic
invariance with its relaxation via collisions \citep[cf.][]{Schekochihin2005}:
\begin{equation}
\Delta{p} = p_\perp-p_\parallel = 0.960\,\frac{p_{\rm i}}{\nu_{\rm ii}}\frac{d}{dt}\ln{\frac{B^3}{\rho^2}},
\label{eqn:pressure_anisotropy}
\end{equation}
where $p_{\rm i}$ is the thermal pressure of the ions and $\nu_{\rm ii}$ is the ion-ion collision frequency. Using
Equations \ref{eqn:density} and \ref{eqn:bfield} to express the rates of change of density and magnetic-field strength
in terms of velocity gradients, and noting that Equation \ref{eqn:spitzer_visc} can be written as $\mu = 0.960p_{\rm i}\nu_{\rm
ii}^{-1}$, Equation \ref{eqn:pressure_anisotropy} becomes 
\begin{equation}
     \Delta{p} = 3\mu\left(\bhat\bhat-\frac{1}{3}\identity\right):\nabla{\bf v} = 3\mu{S},
     \label{eqn:pressure_anisotropy2}
\end{equation}
where we have defined the parallel rate of strain $S = (\bhat\bhat-\identity/3):\nabla{\bf v}$. Substituting this expression for $\Delta{p}$ into Equation \ref{eqn:visc_stress} reproduces Equation \ref{eqn:visc_tensor_braginskii}. In other words, the pressure anisotropy in the plasma is responsible for the Braginskii (anisotropic) viscous stress.

We can also re-arrange the terms in Equation \ref{eqn:stress_tensor} into a part that modifies the total-pressure term and a part that modifies the magnetic-tension term:
\begin{equation}
\ptensor = \left(p+\frac{1}{3}\Delta{p}\right)\identity - \left(\Delta{p} + \frac{B^2}{4\pi}\right)\bhat\bhat.\label{eqn:stress_tensor2}
\end{equation}
As we noted in Section \ref{sec:intro}, when the pressure anisotropy becomes comparable in magnitude to the magnetic
tension, effects come into play that require a kinetic description of the plasma. One can see immediately from the term
proportional to $\bhat\bhat$ in Equation \ref{eqn:stress_tensor2} that when $\Delta{p} \approx -B^2/4\pi$, the magnetic
tension responsible for propagating Alfv\'en waves vanishes, and if $\Delta{p}$ becomes more negative the effective
tension becomes negative. When this occurs, the Alfv\'enic fluctuations grow exponentially, as the plasma becomes
subject to the firehose instability \citep{Rosenbluth1956,Chandrasekhar1958,Parker1958}. The fastest-growing modes of
this instability are near the Larmor scale. The small-scale magnetic fluctuations that result can efficiently
pitch-angle scatter ions, restoring the pressure anisotropy to the marginally stable value of $\Delta{p} \sim
-B^2/4\pi$ \citep{Kunz2014}. 

On the other hand, when $\Delta{p} \gtrsim B^2/8\pi$, the plasma is susceptible to the mirror instability
\citep{Barnes1966,Hasegawa1969,Southwood1993,Hellinger2007}, in which ``magnetic bottle'' configurations (where, in the nonlinear regime, ions
bounce between magnetic mirrors on either side of the bottle) become unstable due to an excess of perpendicular
pressure. In the necks of these bottles where the magnetic-field lines are compressed, the plasma is ``squeezed'' out
due to adiabatic invariance, which lowers the thermal pressure and causes the magnetic-field lines to contract further,
reinforcing the instability. Similarly to the firehose instability, the mirror instability drives the fast growth
of fluctuations near the Larmor scale that, when they become strongly nonlinear, can efficiently pitch-angle scatter ions, restoring the pressure anisotropy to
the marginally stable
value of $\Delta{p} \approx B^2/8\pi$ \citep{Kunz2014,Melville2016}. With both of these instabilities operating, the pressure anisotropy is therefore
constrained to the range \citep{Matteini2006,Hellinger2008,Kunz2014,Riquelme2015}:
\begin{equation}
\nonumber -\frac{B^2}{4\pi} \lesssim \Delta{p} \lesssim \frac{B^2}{8\pi}\label{eqn:limiters},~\rm{or}
 -2\beta^{-1} \lesssim \delta_p \lesssim \beta^{-1},
\end{equation}
where we have defined $\delta_p \equiv \Delta{p}/p_{\rm th}$. The
latter equation illustrates clearly that the limiting of the pressure anisotropy (and, therefore, by Equation \ref{eqn:visc_stress},
of the momentum transport by viscosity) can be an important effect in a high-$\beta$ plasma such as the ICM.

Subsequent work has investigated how other instabilities may play a role in limiting the pressure anisotropy. Strictly
speaking, the firehose instability mentioned above is the ``parallel'' firehose instability, because its fastest-growing unstable fluctuations have wave vectors that are approximately parallel to the local magnetic field's direction. There is also an ``oblique'' firehose
instability, which we consider as well because it drives Alfv\'enic fluctuations unstable before the parallel firehose threshold is reached~\citep{Hellinger2000,Hellinger2008,Bott2021}. As its name suggests, it occurs for wave vectors that are oblique to the local magnetic field direction, in which case the marginal-stability threshold turns out to be $\delta_p
\approx -1.4/\beta$. The numerical factor ($\approx1.4$) does in fact have a weak dependence on $\beta$~\citep{Bott2025}, but for our purposes it suffices to assume that it is constant. Notably, this criterion for marginal stability does not zero out the magnetic tension as the
parallel firehose instability would have done. Another instability that can limit the pressure anisotropy, in the
positive direction in this case, is the ion-cyclotron instability \citep{Gary1997}. This occurs when waves in the plasma
resonantly exchange energy with the ions at their gyrofrequency in the presence of a positive pressure anisotropy, which
can continuously transfer more energy to wave. The marginal stability threshold for this instability is approximately
$\delta_p \approx 0.5/\beta^{0.5}$ \citep{Gary1997,Ley2024}. We will consider all four marginal stability thresholds in our
simulations. 

%
% Subsection 2.2: Code
%
\subsection{Code}\label{sec:code}

We performed our simulations using \code{FLASH} 4.7.1, a parallel conservative magnetohydrodynamic (MHD) astrophysical
simulation code \citep{Fryxell2000,Dubey2009}. The MHD algorithms in \code{FLASH} are detailed in \citet{Lee2013}. The
directionally unsplit corner transport upwind (CTU) integration method and the HLLD Riemann solver are used in all of
our simulations, with third-order piecewise-parabolic \citep[][]{Colella1984} reconstruction. 

We included the Braginskii viscosity following the approaches of \citet{Sharma2007}, \citet{Dong2009},
\citet{Parrish2012}, \citet{Kunz2012}, \citet{ZuHone2015}, and \citet{Kingsland2019}. The viscous momentum fluxes are
implemented via operator splitting, with monotonized central (MC) limiters applied to the fluxes to preserve
monotonicity. The latter ensures that unphysical transport of energy and momentum does not occur in the presence of
steep gradients. Since these diffusive processes are modeled by explicit time-stepping methods, they have very
restrictive Courant-limited timesteps $\propto (\Delta{x})^2$. Finally, since we are only concerned in this work with
the effects of viscosity on the properties of the sloshing gas (including the bulk motions and turbulence which result),
we do not explicitly include the effects of radiative cooling during the simulation, though our initial condition
implicitly assumes that the cool core was originally formed by such cooling.

In order to test the effects of different pressure-anisotropy limiting prescriptions due to the effects of plasma
instabilities, we implemented three such prescriptions when calculating the viscous flux at every timestep. The first,
which we will refer to as ``Unlimited,'' does not apply any limits to the pressure anisotropy. It should be noted that
the Braginskii-MHD equations are ill-posed, given that the aforementioned plasma instabilities arise from a kinetic
description and are not captured by these equations (in this case, solutions to these equations are regularized by the
finite resolution of the grid). Nevertheless, we include these simulations to explore the limit of an unsuppressed
Braginskii viscosity. The second, ``Limiters v1,'' restricts the pressure anisotropy to the range $-1.4\beta^{-1} \leq
\delta_p \leq \beta^{-1}$, assuming that the mirror and oblique firehose marginal stability thresholds provide the
limits on the pressure anisotropy. The third, which we will refer to as ``Limiters v2,'' restricts the pressure
anisotropy to the range $-2\beta^{-1} \leq \delta_p \leq 0.53\beta^{-0.4}$ \citep{Ley2024}, employing the
least restrictive parallel firehose and ion-cyclotron marginal-stability thresholds. 

While the conditions in the ICM are such that the mirror instability is expected to be the main instability that limits
the growth of positive pressure anisotropies, the ion-cyclotron instability is also expected to be relevant as a
secondary instability that is excited by the mirror instability itself \citep{Ley2024}. Additionally, there is evidence
of the appearance of the ion-cyclotron instability alongside the mirror instability in high-beta kinetic turbulence
simulations \citet{Arzamasskiy2023}. This is why we use the ion-cyclotron-instability threshold in ``Limiters v2.'' In
all of the cases, we implement the limits on the pressure anisotropy as ``hard-wall'' limiters---any value of
$\Delta{p}$ arising from Equation \ref{eqn:pressure_anisotropy} that exceeds the limits is set to the limit value. Such
limiters can be applied when the effective scattering rates anticipated from these instabilities \citep[$\sim \beta
(\bhat\bhat - \identity/3):\nabla{\bf v}$, see][]{Kunz2014}, while large compared to fluid-dynamical evolution rates,
are much smaller the ion gyrofrequency $\Omega_{\rm i}$, which for the thermodynamic and kinematic conditions in the ICM
is satisfied if the magnetic field strength $B \gg 1$~nG \citep{StOnge2020}. This is safely satisfied in our simulations
of a developed cool-core cluster.  

\subsection{Initial Conditions}\label{sec:ICs}

The merger scenario that we use in this work is identical to our previous setup in \citet{ZuHone2011a}, which was
originally derived from \citet{AM06}. In this case, the cool-core cluster is more massive ($M \sim 10^{15}~{\rm
M}_\odot$) and hotter ($T \sim 7$~keV), resembling A2029 (though not exactly reproducing it). The gas-less subcluster is
5 times less massive than the main cluster and approaches it with an initial impact parameter of $b = 500$~kpc.

The gravitational potential on the grid is the sum of two collisionless ``rigid bodies'' corresponding to the
contributions to the potential from both clusters. This approach to modeling the potential is used for simplicity and
speed over solving the Poisson equation for the matter distribution, and is an adequate approximation for our purposes.
It is the same approach that we used in previous works, and is justified and explained in \citet{Roediger2012}. 

\begin{figure*}[!th]
\centering
\includegraphics[width=0.98\textwidth]{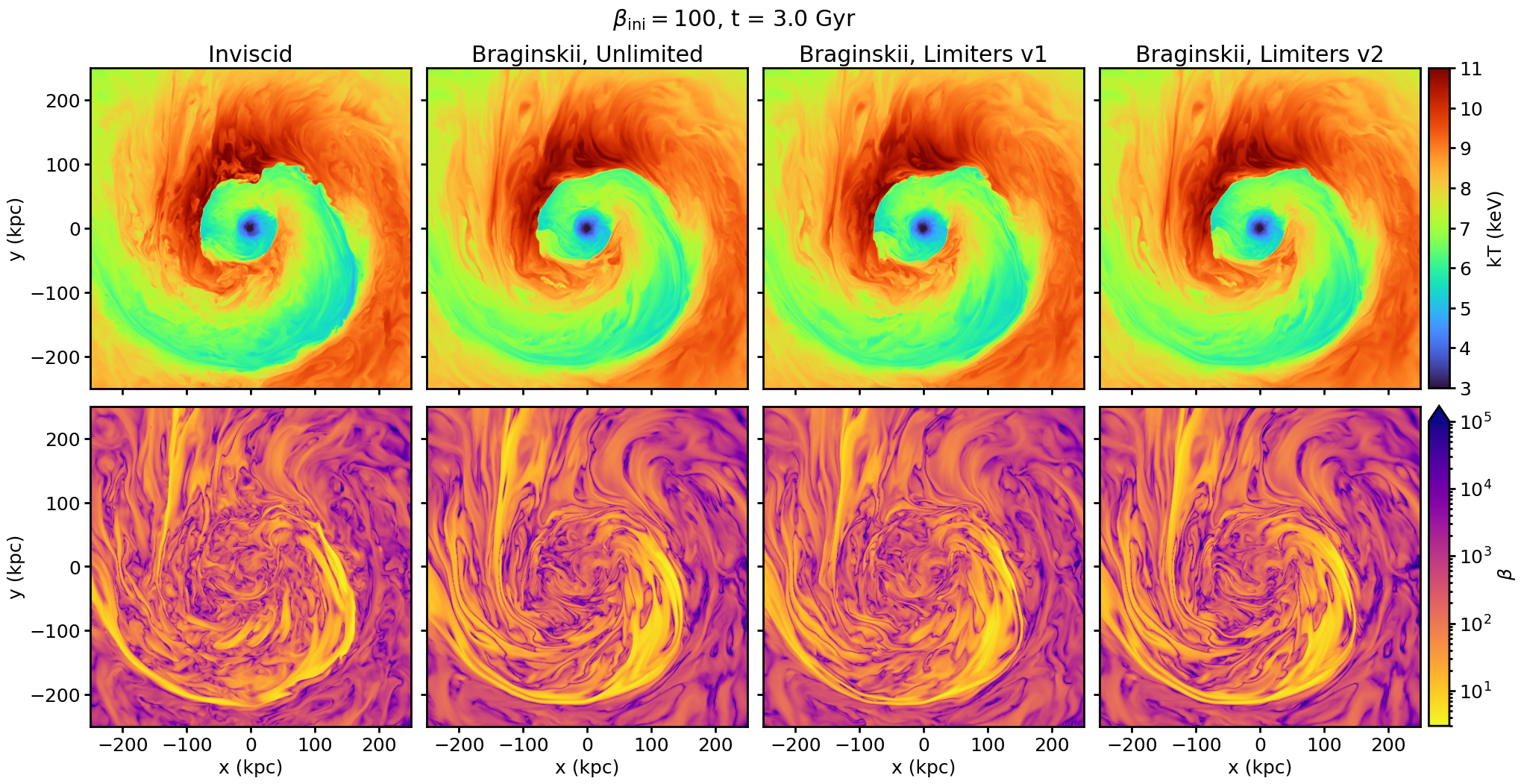}
\includegraphics[width=0.98\textwidth]{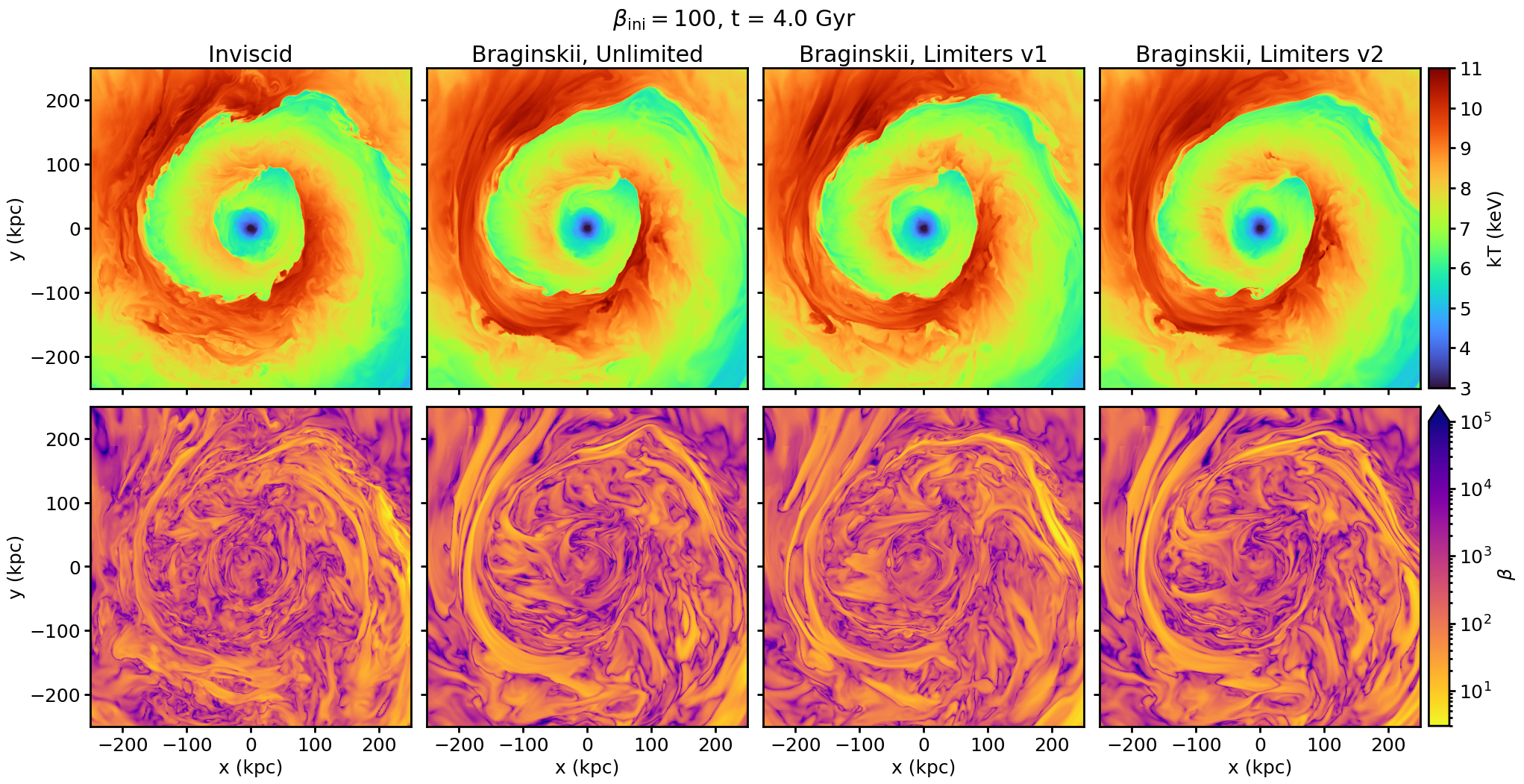}
\caption{Slices in the $x-y$ plane (through the center of the domain) of the gas temperature $kT$ (top panels) and the plasma $\beta$ (bottom panels) for the simulations with $\beta_{\rm ini}$ = 100, at the epochs $t$ = 3.0 and 4.0~Gyr. Each panel is 500~kpc on a side.\label{fig:beta100_temp_beta}}
\end{figure*}

The tangled magnetic field of the cluster in the initial conditions is also set up in a similar way to previous works.
The average magnetic-field strength at a radius $r$ from the cluster center is set by enforcing a constant ratio
$\beta_{\rm ini} = p_{\rm th}/p_B$. In this work, we have two sets of simulations with two different values of
$\beta_{\rm ini}$: 100 and 400. These two different values allow us to explore the effect of different magnitudes of
magnetic tension in comparison to Braginskii viscosity in the different simulations. For the three-dimensional
tangled structure of the initial magnetic field, we follow the approach of \citet{ZuHone2011a} and \citet{ZuHone2015},
and we refer the reader to those papers for the details. 

As detailed in Appendix A of \citet{ZuHone2011a}, the initially tangled magnetic field is not an equilibrium field.
Before the passage of the subcluster, the field slowly begins to relax to a new configuration, transferring some of its
energy to the gas in the form of low-velocity turbulent motions, which do not significantly affect the overall
thermodynamic state of the gas. However, the important result for this work is that the plasma $\beta$ at later times becomes larger than the initial value, corresponding to a decrease in the average magnetic-field strength. In Appendix \ref{sec:magnetic_relaxation}, we show the extent of this evolution before the subcluster passage. Importantly, the main results of this work are not significantly impacted by this effect.

For all of the simulations, we set up the main cluster within a cubical computational domain of width $L = 2$~Mpc on a
side. We employ adaptive mesh refinement (AMR), with a base grid of 128 cells on a side, and 5 levels of refinement,
resulting in a finest cell size of $\Delta{x} = 0.98$~kpc. The only refinement criterion is that the cells within a
radius of $r = 250$~kpc from the main cluster's potential minimum (and the center of the domain) be maximally refined.

Similar to the approach taken in \citet{ZuHone2015}, in  simulations with viscosity, we enable the viscous stresses at
the time $t$ = 1.66~Gyr, shortly after the pericenter passage. This is just before the onset of the sloshing motions,
and allows us to examine the effects of viscosity during the period of most significant gas motions, without viscosity
affecting the conditions of the cluster in the time before the pericenter passage, to allow for a more precisely
controlled experiment. 

% Section 3
%
\section{Results}\label{sec:results} 
          
\subsection{Properties of the Temperature and Magnetic Field}\label{sec:slices_temp_beta}

\begin{figure*}
\centering
\includegraphics[width=0.98\textwidth]{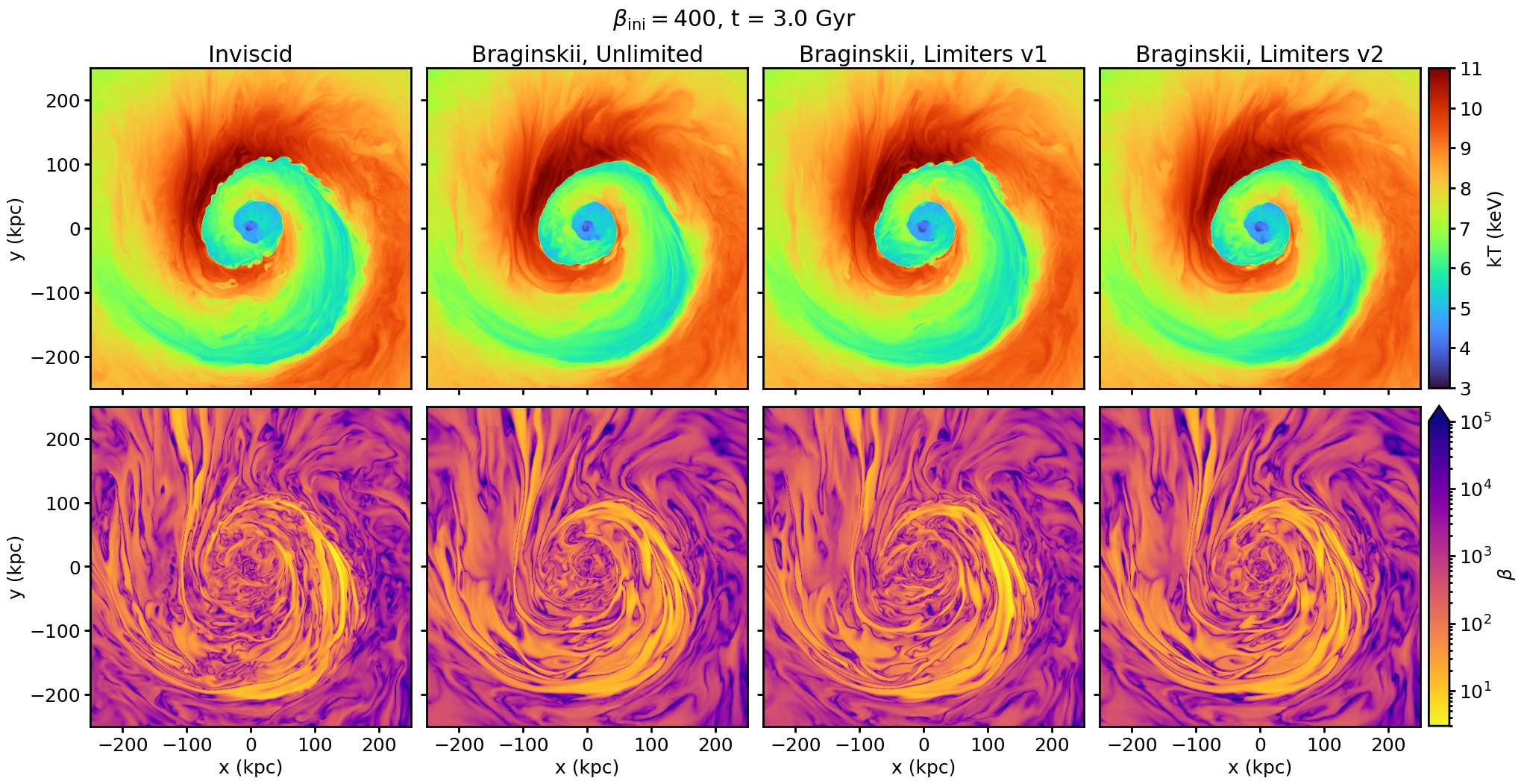}
\includegraphics[width=0.98\textwidth]{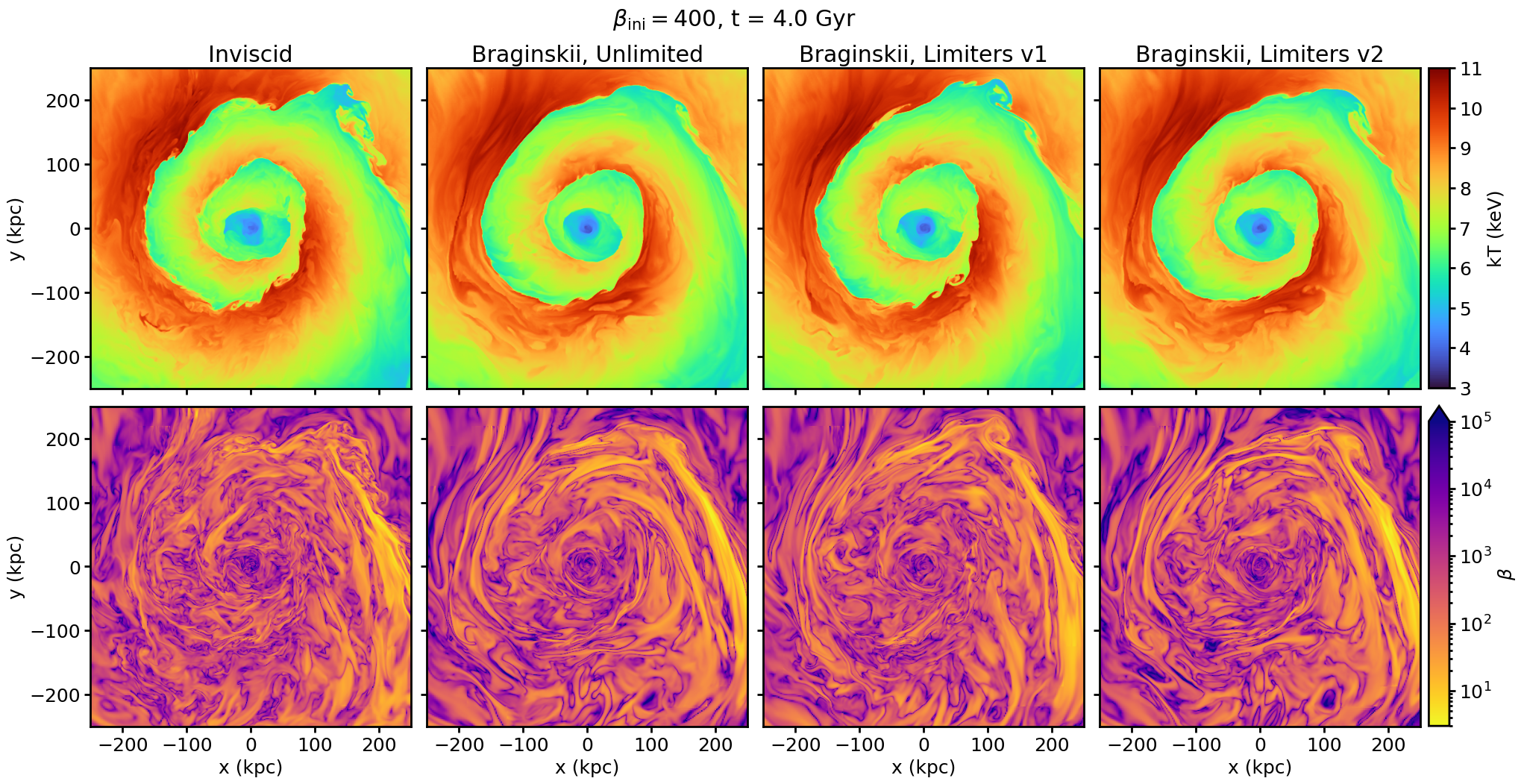}
\caption{Same as Figure \ref{fig:beta100_temp_beta}, but for the simulations with $\beta_{\rm ini}$ = 400.\label{fig:beta400_temp_beta}}
\end{figure*}

\begin{figure*}[!ht]
\centering
\includegraphics[width=0.48\textwidth]{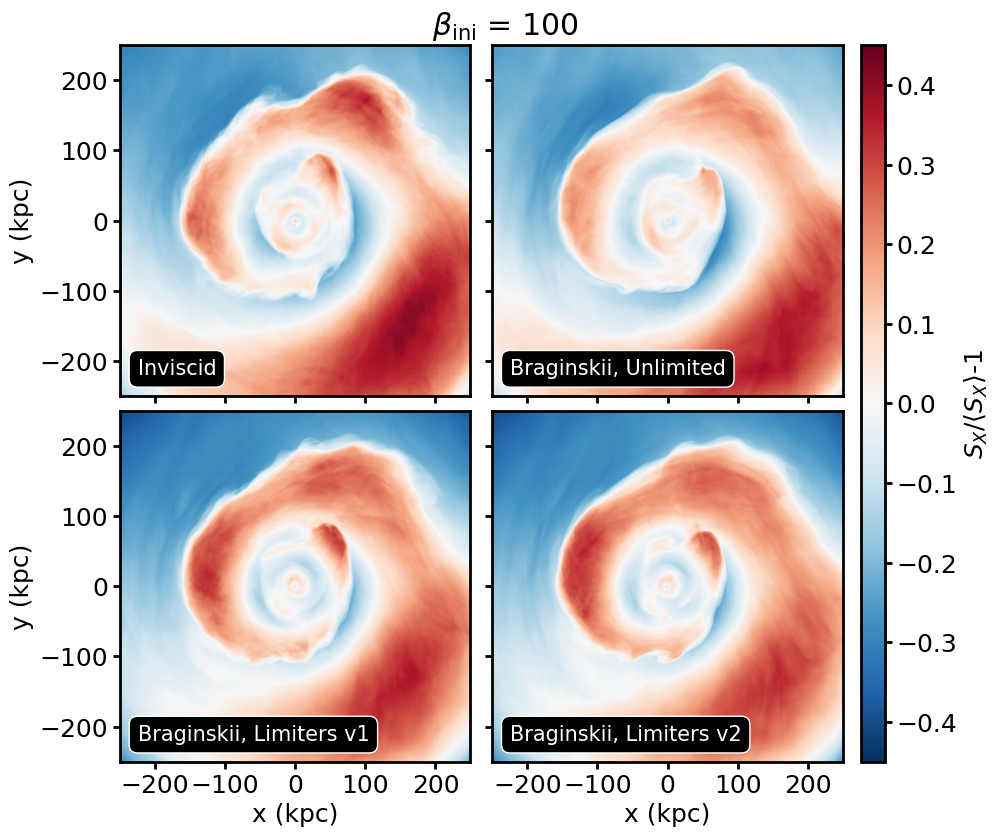}
\includegraphics[width=0.48\textwidth]{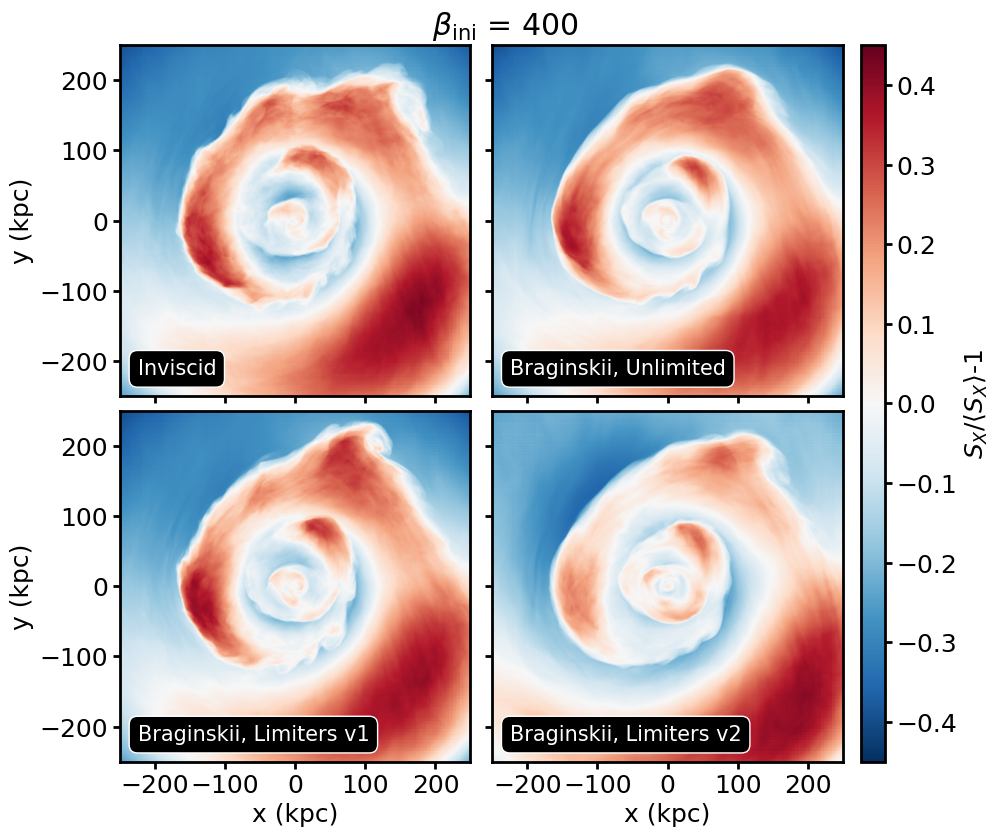}
\caption{Residual maps of the X-ray SB $S_X$ in the 0.5-4~keV band, obtained by fitting the SB image to a best-fit axisymmetric model composed of a sum of two $\beta$-models and computing $S_X/\langle{S_X}\rangle - 1$. All simulations are shown ($\beta_{\rm ini}$ = 100 on the left and $\beta_{\rm ini}$ = 400 on the right) at $t$ = 4.0~Gyr. The simulation was projected along the $z$-axis, perpendicular to the main sloshing plane.\label{fig:resid_maps}}
\end{figure*}

\begin{figure*}[!ht]
\centering
\includegraphics[width=0.96\textwidth]{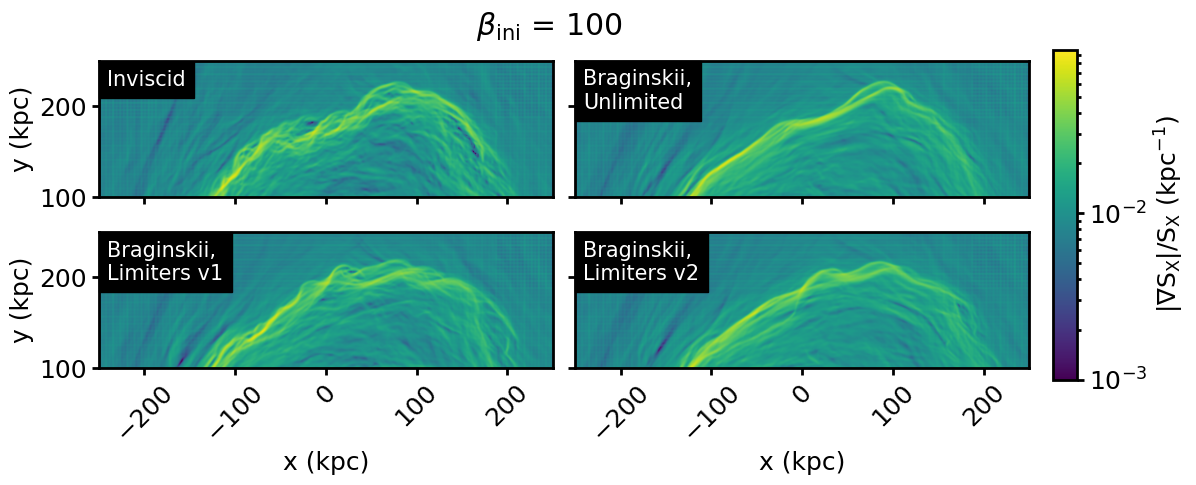}
\includegraphics[width=0.96\textwidth]{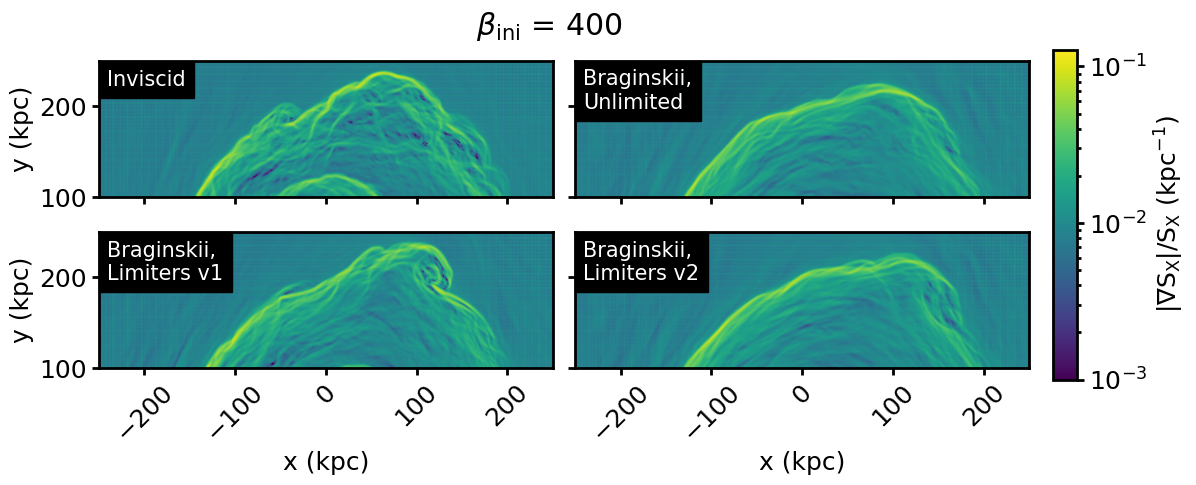}
\caption{Gaussian gradient magnitude (GGM) images of the projected X-ray SB $S_X$ of the northern CF in the 0.5-7.0~keV band for all of the simulations ($\beta_{\rm ini}$ = 100 on the top and $\beta_{\rm ini}$ = 400 on the bottom), at $t$ = 4.0~Gyr. The simulation was projected along the $z$-axis, perpendicular to the main sloshing plane.\label{fig:ggm_maps}}
\end{figure*}

We begin by showing slices of the gas temperature and plasma $\beta$ in the merger plane for all eight simulations, at
the two epochs of $t$ = 3.0 and 4.0~Gyr. Throughout this work, we will focus on these epochs because they
capture the key dynamics of the developed sloshing motions and the effects on the turbulence, magnetic field, and
pressure anisotropy that result from them. There is little qualitative difference between these two epochs for most of
the results presented here, so we will usually present only one or the other for the sake of brevity. Figure
\ref{fig:beta100_temp_beta} shows these for the simulations with $\beta_{\rm ini}$ = 100. In each, the left-most panels
show the inviscid simulation and the rest of the panels show simulations with Braginskii viscosity using the three
different limiting prescriptions for the pressure anisotropy. The main characteristics of these slices are the same as
shown in previous studies, such as \citet{ZuHone2011a,ZuHone2015}: sloshing gas motions produce spiral-shaped CFs and in
the regions dominated by the sloshing motions, the magnetic field is amplified to $\beta \lesssim 10$ in thin structures
largely (though not always) aligned with the CF surfaces. The magnetic field varies more in the direction normal to the
CFs than in the tangential directions, due to the fact that the super-Alfv\'enic sloshing motions have stretched out
most of the significant fluctuations in the latter direction.
     
\begin{figure*}[!t]
\centering
\includegraphics[width=0.90\textwidth]{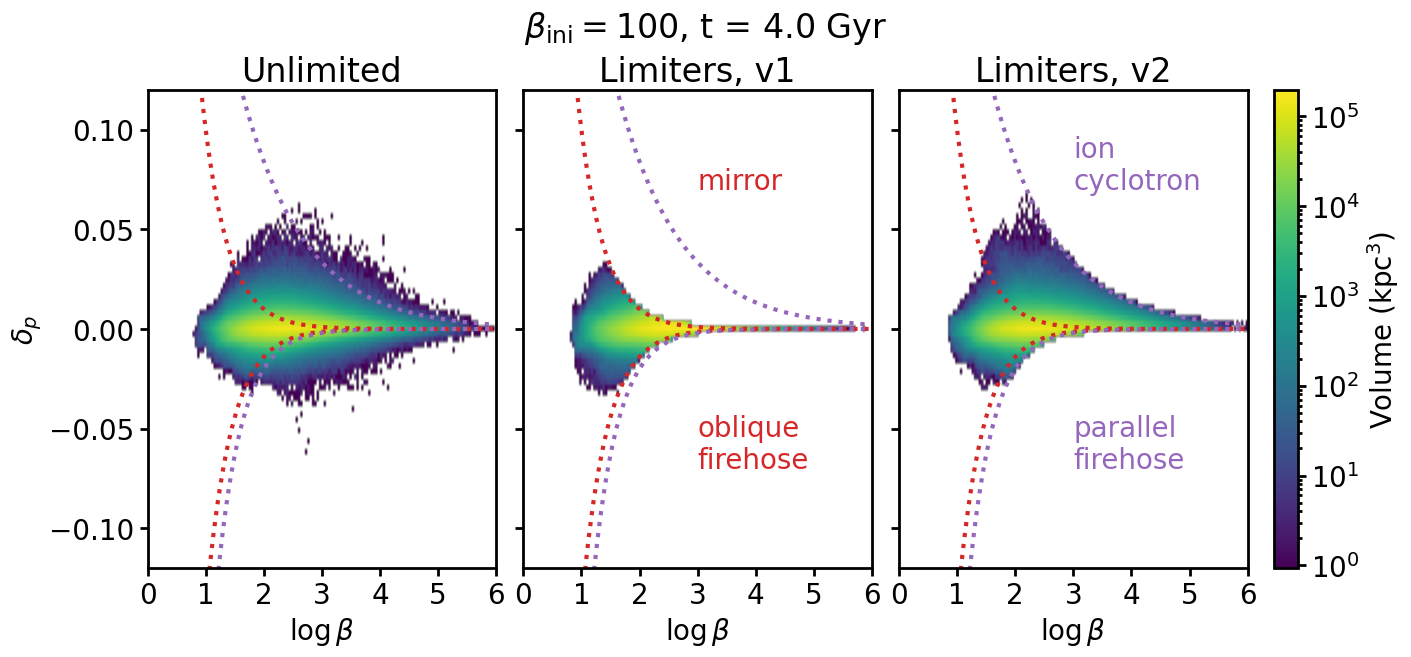}
\includegraphics[width=0.90\textwidth]{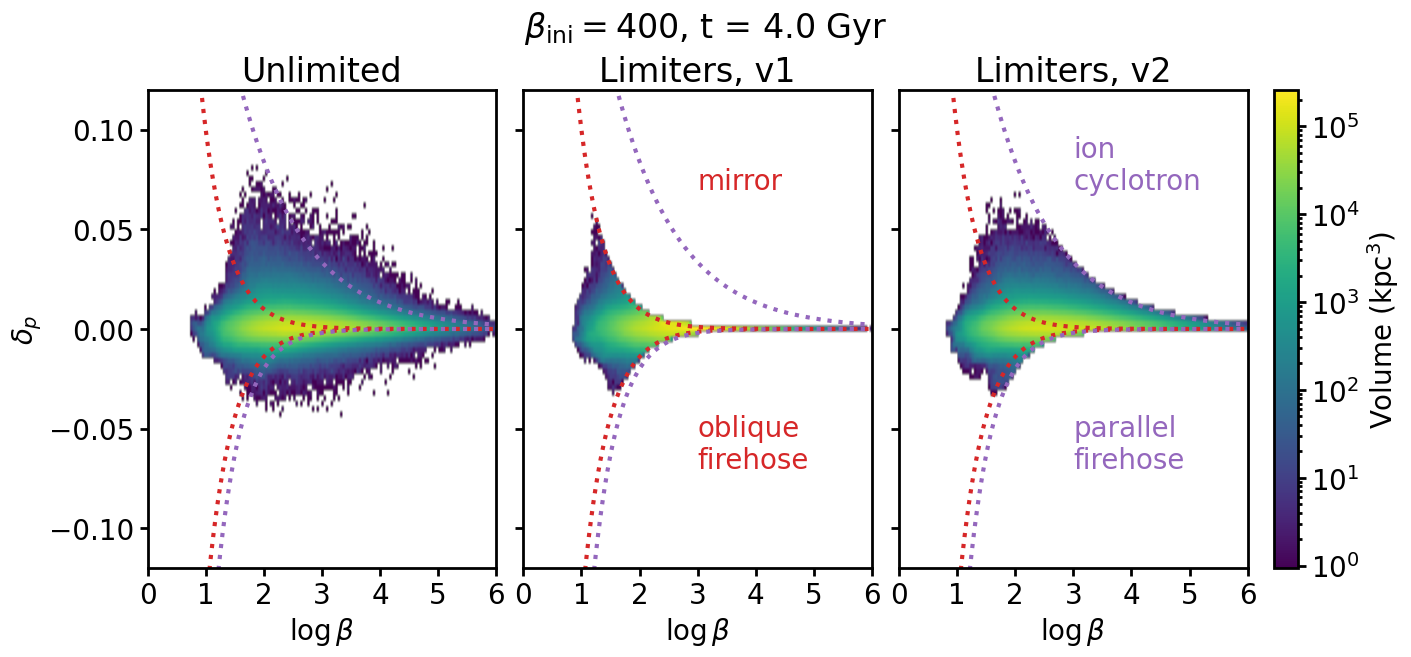}
\caption{Parameter-space plots of the fractional pressure anisotropy $\delta_p$ vs. the plasma $\beta$ for the
$\beta_{\rm ini} = 100$ and $\beta_{\rm ini} = 400$ simulations at $t = 4.0$~Gyr. The colormap represents the total cell
volume at each ($\delta_p$, $\beta$) value pair. The dashed lines indicate the bounds imposed by the two different
limiter schemes.\label{fig:beta_phase}}
\end{figure*}

\begin{figure*}[!t]
\centering
\includegraphics[width=0.48\textwidth]{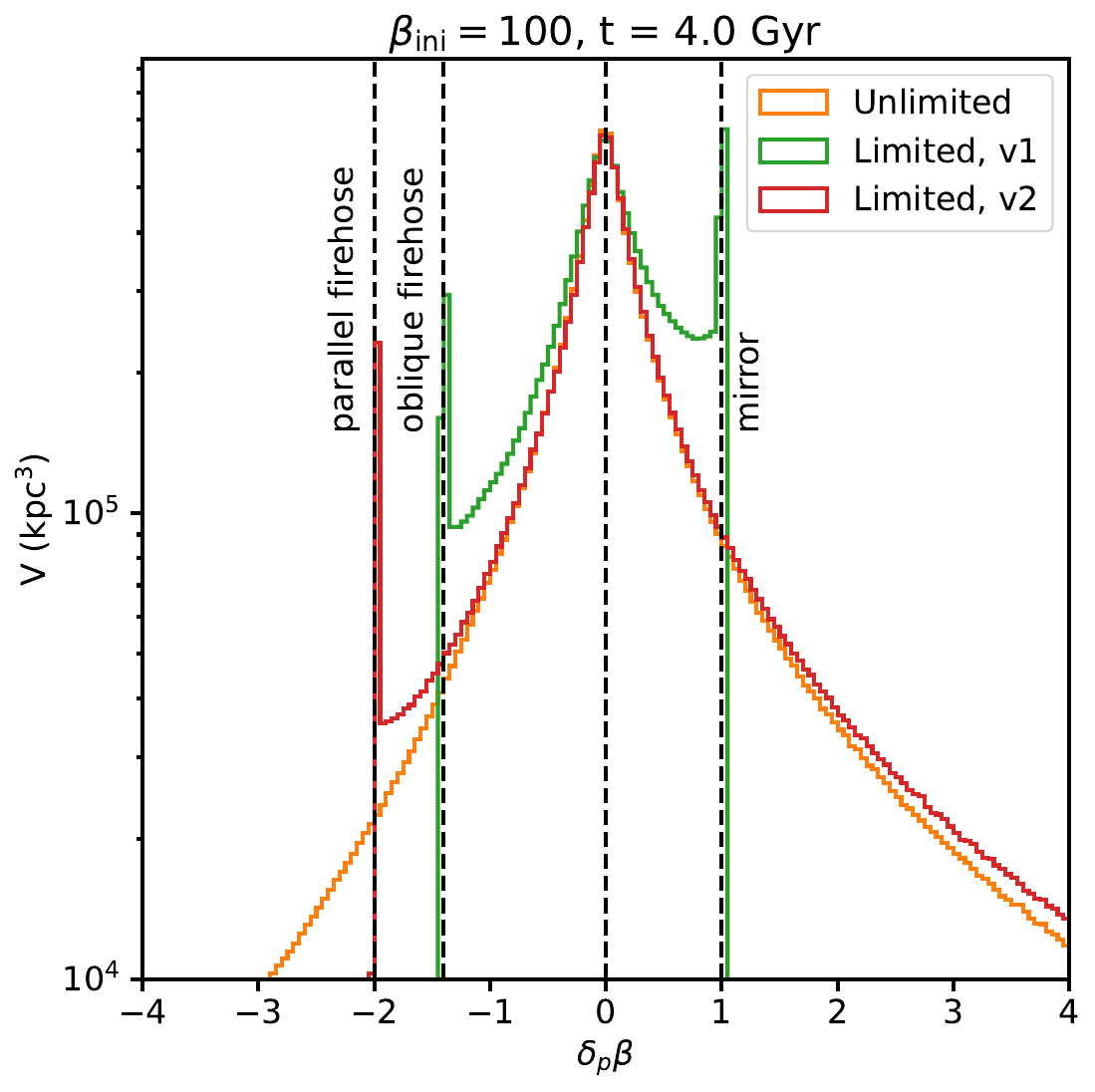}
\includegraphics[width=0.48\textwidth]{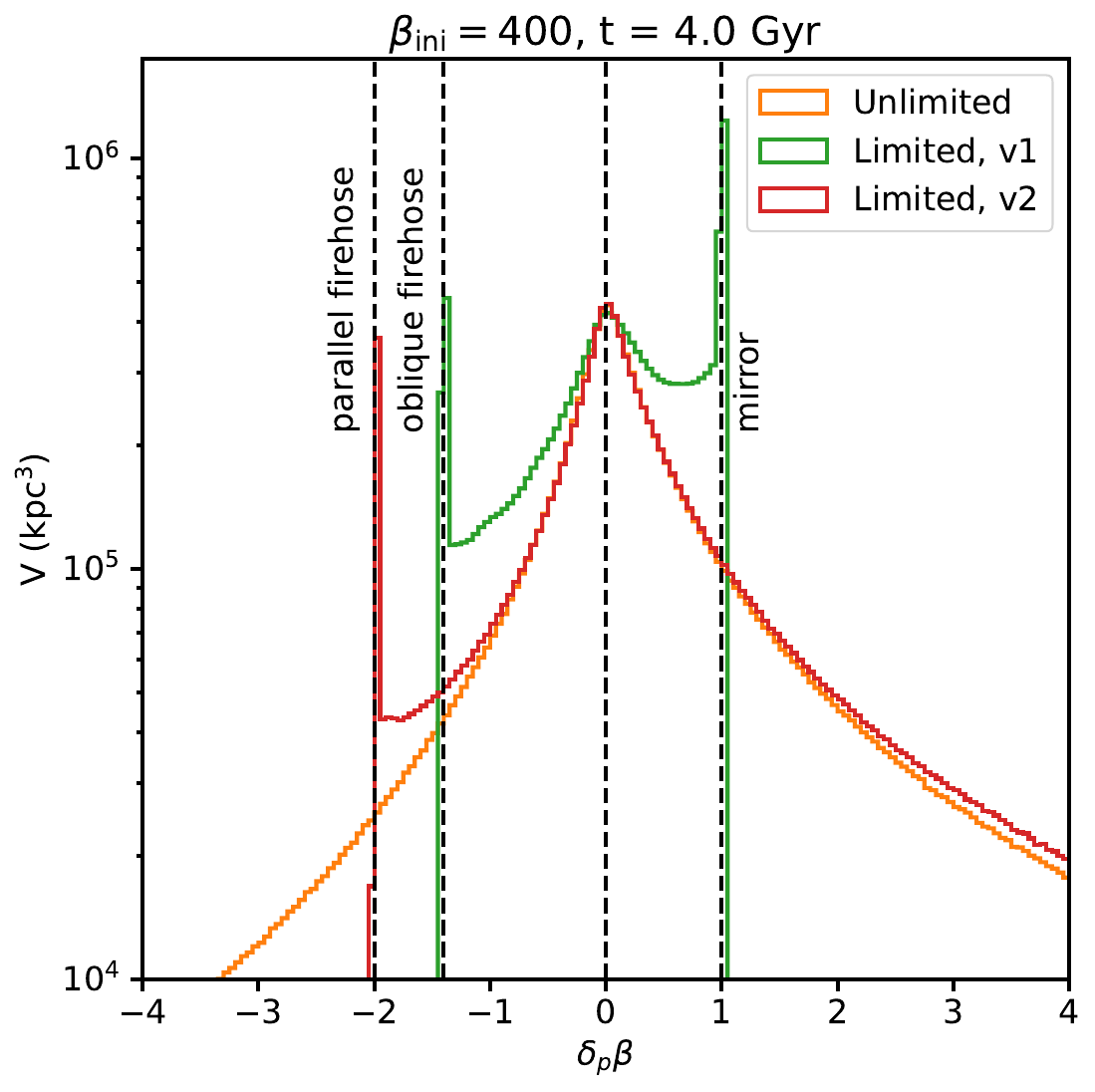}
\caption{Histograms of $\delta_p\beta$ for the $\beta_{\rm ini} = 100$ and $\beta_{\rm ini} = 400$
simulations at $t = 4.0$~Gyr. The dashed lines indicate the bounds imposed by the two different limiter
schemes.\label{fig:delta_beta_hist}}
\end{figure*}

\begin{figure*}[!t]
\centering
\includegraphics[width=0.98\textwidth]{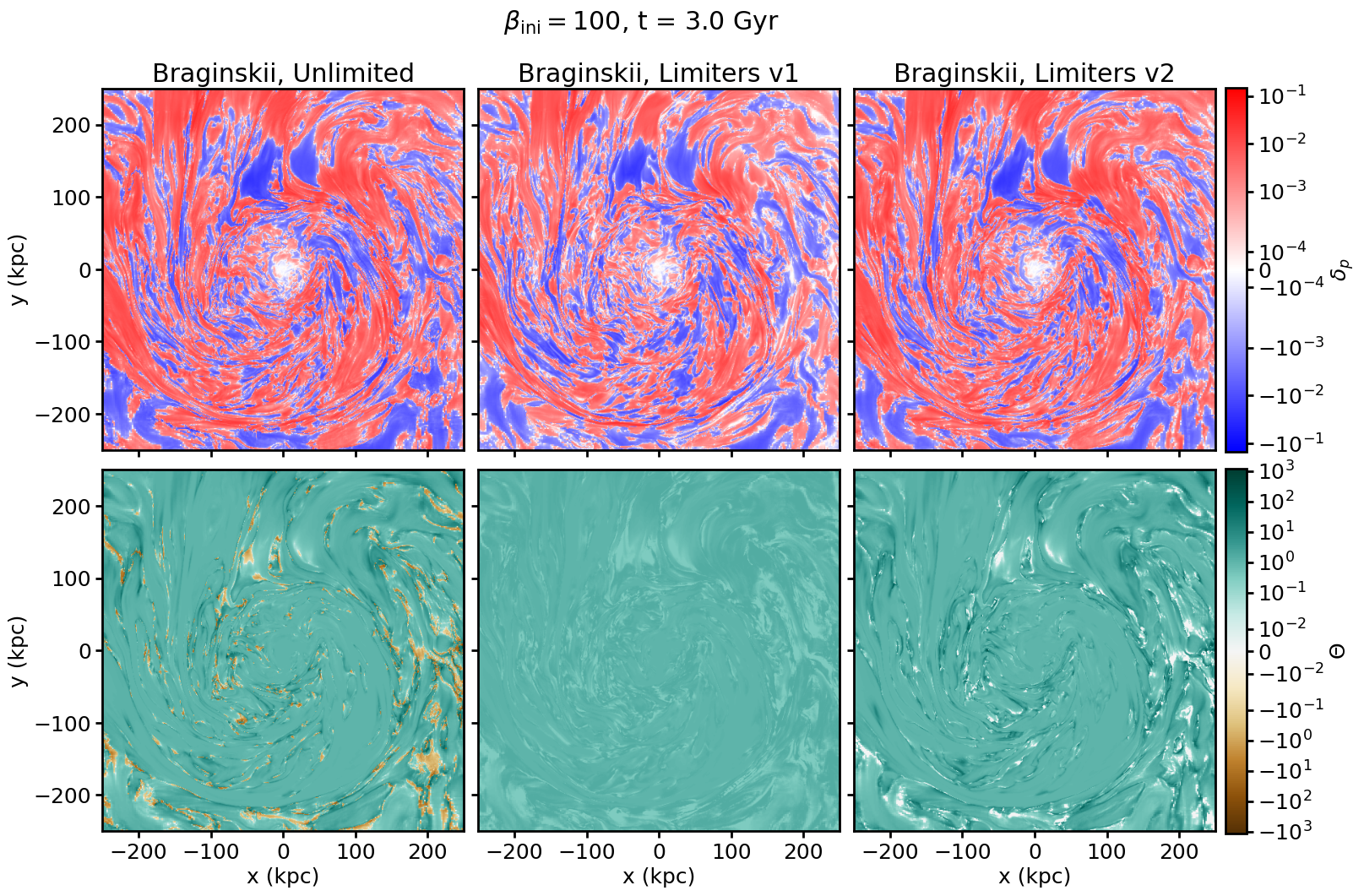}
\includegraphics[width=0.98\textwidth]{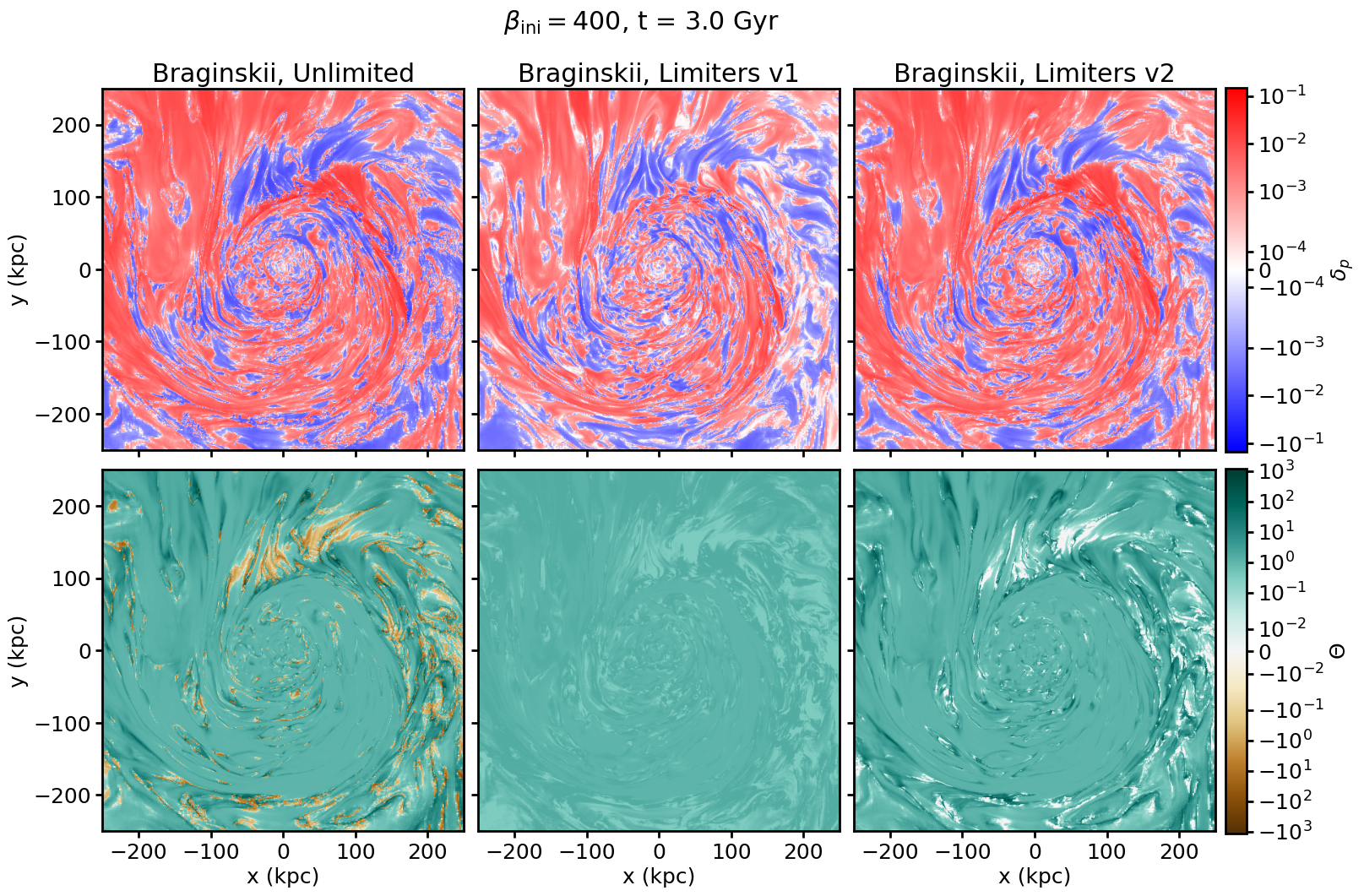}
\caption{Slices through the center of the domain showing $\delta_p$ (first and third rows) and $\Theta$ (second and fourth rows; defined in Equation \ref{eqn:theta}) for the $\beta_{\rm ini} = 100$ (top six panels) and $\beta_{\rm ini} = 400$ (bottom six panels) simulations at $t = 3.0$~Gyr.\label{fig:delta_ratio}}
\end{figure*}

\begin{figure*}
\centering
\includegraphics[width=0.96\textwidth]{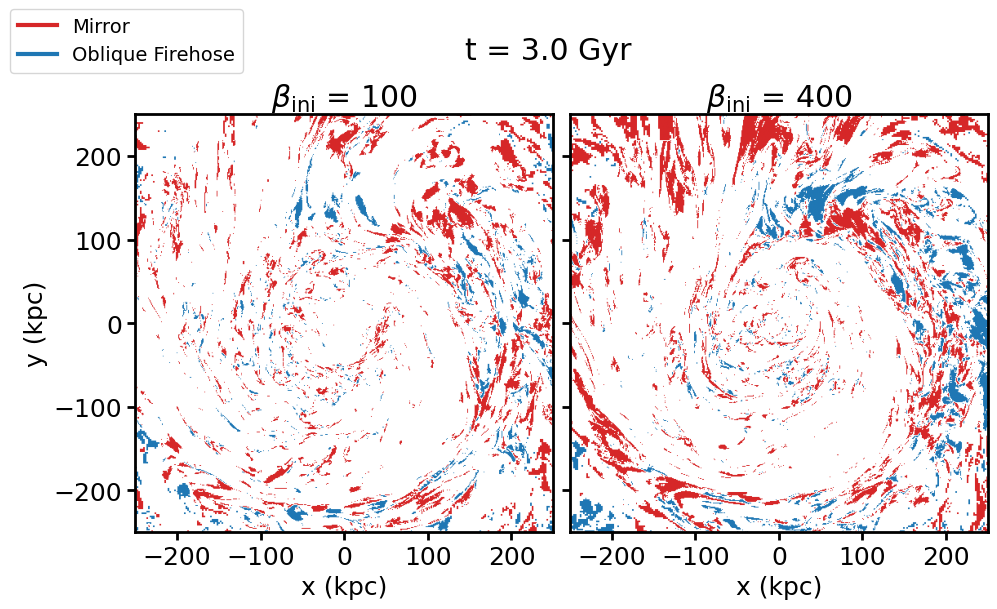}
\caption{Slices through the center of the domain at $t = 3.0$~Gyr, showing for both values of $\beta_{\rm ini}$, the locations in the ``Limiters, v1'' simulations where the pressure anisotropy takes the values of the mirror (red) and oblique-firehose (blue) instability limits.\label{fig:limiter_maps}}
\end{figure*}

The addition of Braginskii viscosity results in noticeable changes in the appearance of the CFs in temperature and
magnetic-field structure. In the inviscid simulations, the CFs are noticeably distorted by KHI, which manifest as
wave-like features on the surface (most obvious in temperature, top panels). In the slices of $\beta$, this is
manifested as a more tangled and disordered magnetic field. When ``Unlimited'' Braginskii viscosity is
included (left-center panels), the KHI are suppressed, resulting in smoother CFs and more laminar magnetic field
structures. When the pressure anisotropy is limited (right-center and right-most panels), the KHI are still suppressed,
but not quite as effectively as in the ``Unlimited'' case, especially in the ``Limiters, v1'' case, which has the more
restrictive limits on pressure anisotropy.

Similarly, Figure \ref{fig:beta400_temp_beta} shows slices of the gas temperature and plasma $\beta$ for the simulations
with $\beta_{\rm ini}$ = 400. Here, the initial magnetic pressure and tension are a factor of 4 lower, the effects of
KHI on the CFs are more significant, and the value of $\beta$ in the thin magnetic field structures is not as low as the
cases with $\beta_{\rm ini}$ = 100 \cite[as already shown by][]{ZuHone2011a}. As a result, viscosity plays a larger role
in suppressing KHI at CF surfaces, as seen in the temperature maps. Also, since the initial (and subsequent) magnetic
field strength is lower, the viscous flux is more limited by the pressure anisotropy limits in the two simulations where
the pressure anisotropy is limited, resulting in slightly less suppression of KHI than in the ``Unlimited'' case.
     
\subsection{Effects on Cold Front Stability as Seen in X-rays}\label{sec:xray_results}

The thermal emission from the ICM is observed in the X-ray band, so evidence for suppression of KHI at CFs from
viscosity may appear in projections of X-ray SB $S_X$, which is proportional to $n_en_H\Lambda(T,Z)$,
where $n_e$ and $n_H$ are the electron and hydrogen number densities, respectively, and $\Lambda(T,Z)$ is the X-ray
emissivity, which depends on the temperature $T$ and abundance $Z$ of the gas. We make projections of the predicted
X-ray SB in our simulations using the Astrophysical Plasma Emission Code \citep[APEC,
v3.1.2][]{Smith2001,Foster2012} model. APEC is an appropriate model for a plasma in collisional ionization equilibrium
such as the ICM. Since our simulations do not include a metallicity field, we assume a uniform metallicity of $Z = 0.3
Z_\odot$ for the ICM. 

Various techniques have been used in observed data to highlight the presence of SB features in X-ray
images clusters. One such technique is to fit the X-ray SB image to a smooth model and then divide the
image by the model to produce a residual map. We show such residual maps for all of our simulations projected along the
$z$-axis at $t$ = 4.0~Gyr in Figure \ref{fig:resid_maps}. The residual maps are produced by fitting the projected X-ray
SB image to a best-fit axisymmetric model composed of a sum of two $\beta$-models, and computing
$S_X/\langle{S_X}\rangle - 1$, where $\langle{S_X}\rangle$ is the fitted model. Where viscosity is absent or most
limited (``Inviscid'' and ``Limiters v1''), fronts appear more disrupted in projection by KHI, in contrast to the two
other simulations. Consistent with the discussion in Section \ref{sec:slices_temp_beta}, these differences are not as
significant in the simulations with $\beta_{\rm ini}$ = 100 because the increased magnetic-field strength in these
simulations results in a more effective suppression of KHI along the CF surfaces (even in the inviscid case) and the
viscous fluxes are not as constrained by the pressure-anisotropy limits. 

Fluctuations in SB on smaller scales can be discerned underneath the CF surfaces as well. Such features in $S_X$ can
also be highlighted in X-ray images using gradient-filtering techniques on X-ray images, such as the Gaussian gradient
magnitude (GGM) \citep{Sanders2016}, in which an image is smoothed by a Gaussian and then the gradient is taken. Figure
\ref{fig:ggm_maps} shows the GGM of the projected X-ray SB images divided by the SB itself ($|\nabla{S_X}|/{S_X}$) in
the vicinity of the northern CF for the simulations with $\beta_{\rm ini}$ = 100 (top) and $\beta_{\rm ini}$ = 400
(bottom), at $t$ = 4.0~Gyr, using a standard deviation of 1~kpc for the Gaussian smoothing kernel. These images are very
similar to Figure 8 from \cite{Roediger2013}, which shows Sobel-filtered images of X-ray emission from a CF in
simulations of the Virgo Cluster with varying isotropic viscosity. Here, the differences between the simulations are
obvious, not only in terms of the evidence of disruption of the CFs by KHI (as in Figures
\ref{fig:beta100_temp_beta}-\ref{fig:resid_maps}), but also in terms of smaller-scale SB fluctuations
observed underneath the CFs. Where viscosity is absent or most limited (``Inviscid'' and ``Limiters v1''), the
small-scale fluctuations in $S_X$ underneath the CFs are more pronounced. When viscosity is not, or less, limited
(``Braginskii'' and ``Limiters v2''), the $S_X$ fluctuations are suppressed more strongly, and what remains of the
latter are more laminar in character, following the distribution of the magnetic field and varying in the direction
normal to the CF surfaces where the magnetic field strength is varying on smaller length scales. We will examine the
properties of SB fluctuations in more detail in Section \ref{sec:v_from_sb}.

\begin{figure*}[!t]
\centering
\includegraphics[width=0.97\textwidth]{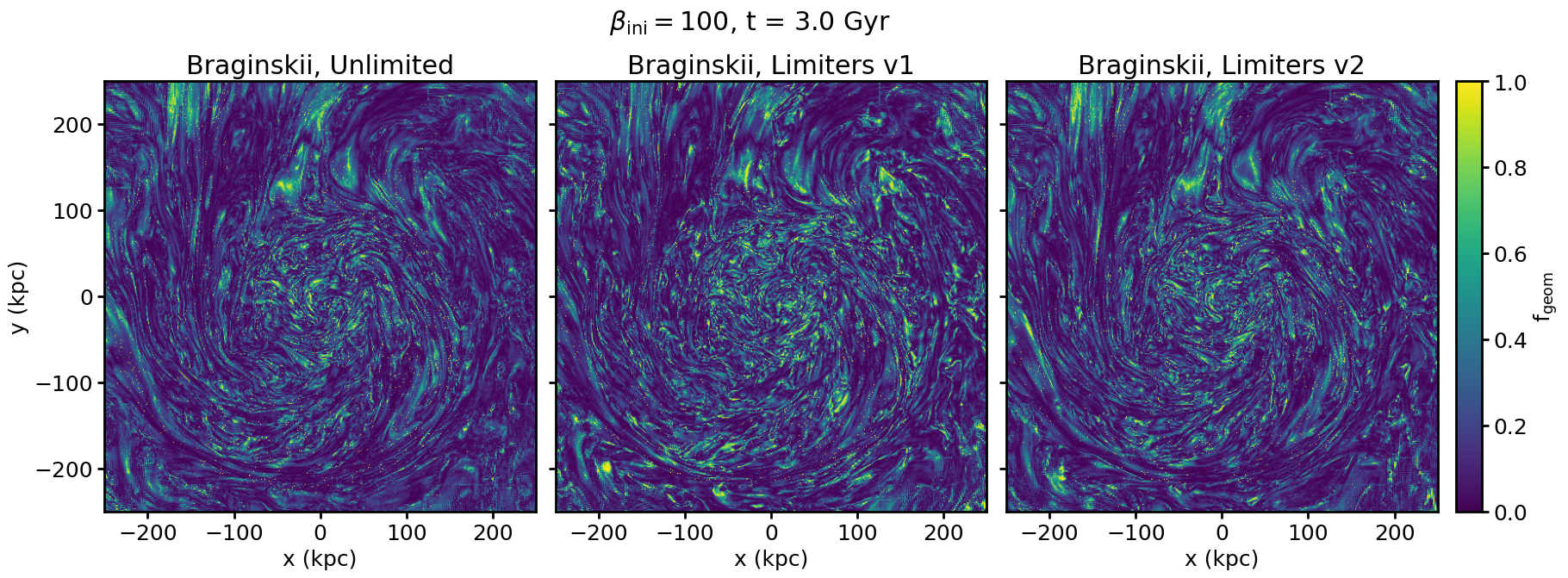}
\includegraphics[width=0.97\textwidth]{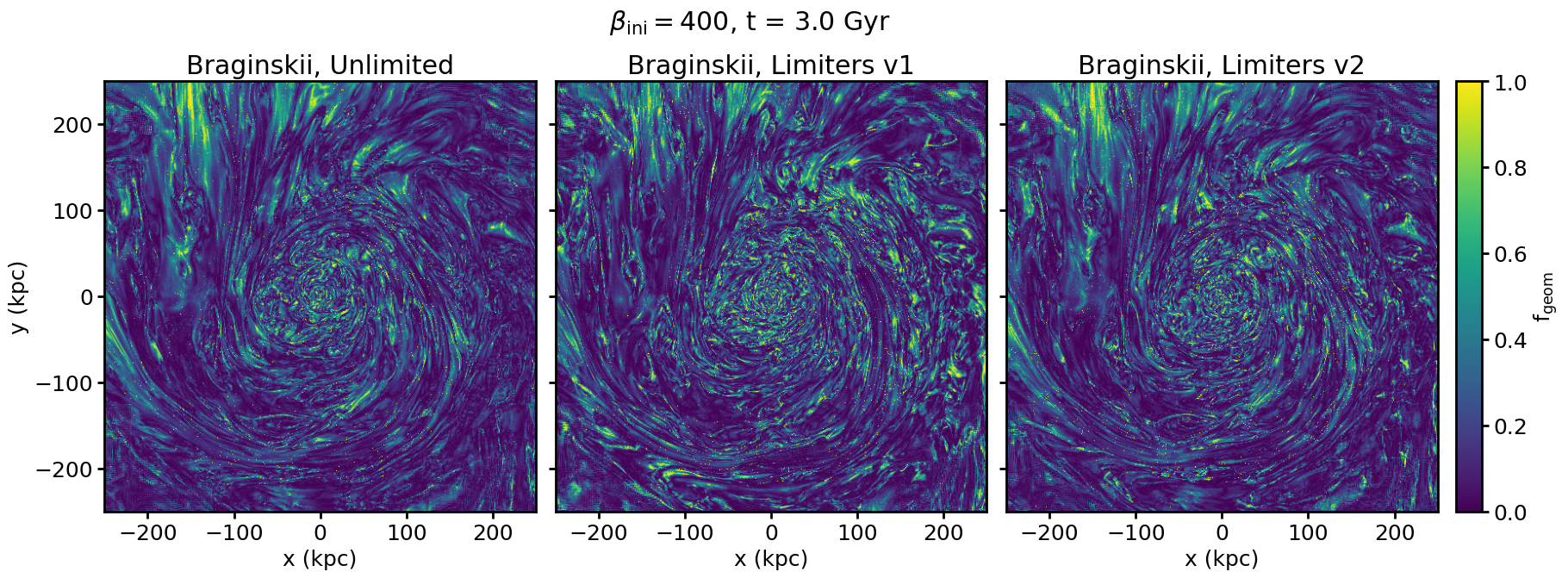}
\caption{Suppression of viscous stress due to the field-line geometry. Panels show slices through $f_{\rm geom}$, defined in Equation \ref{eqn:fgeom}, for the $\beta_{\rm ini} = 100$ and $\beta_{\rm ini} = 400$ simulations at $t$ = 3.0~Gyr.\label{fig:fgeom}}
\end{figure*}

\subsection{Properties of the Pressure Anisotropy}\label{sec:anisotropy_results}

\subsubsection{Phase Space and Histograms of the Pressure Anisotropy}\label{sec:paniso_hist}

We now turn to examining the properties of the pressure anisotropy in the simulations. Figure \ref{fig:beta_phase} shows
phase plots of the pressure anisotropy $\delta_p$ versus the plasma $\beta$ (also known as ``Brazil'' plots) within a
sphere of radius 150~kpc from the cluster center for all the simulations with viscosity, at $t = 4.0$~Gyr (the results
are very similar for other epochs). The colormap indicates the total volume at a given value of $\delta_p$ and $\beta$.
The red and purple lines indicate the boundaries in this space of the ``v1'' and ``v2'' limiters, respectively. In the
``Unlimited'' cases, most of this volume has a pressure anisotropy that falls within the various plasma instability
limits (the yellow color in the phase plots), but there is a fraction of the volume that does have values of $\delta_p$
that fall outside these limits. In the $\beta_{\rm ini}$ = 100 ``Unlimited'' simulation, $\sim$20/8\% of the volume
falls outside the mirror/oblique-firehose limits, whereas for the $\beta_{\rm ini}$ = 400 case the fractions are
slightly larger, $\sim$30\% and $\sim$11\%, respectively. For the less-restrictive ion-cyclotron and parallel-firehose
limits in both simulations, the fractions are much smaller; less than $\sim$1\% for the former and $\sim$5-8\% for the
latter. In both of the ``Unlimited'' simulations, the maximum absolute values of the pressure anisotropy are $|\delta_p|
\sim 0.03-0.07$ at $\beta \sim 100-1000$. For these portions of the phase space at the extremes of the distribution, in
a small portion of the volume, the corresponding product $\beta|\delta_p| \sim 3-70$ indicates that the pressure
anisotropy in the ``Unlimited'' simulations can reach values significantly larger than the magnetic pressure and
tension.

There is a subtle difference between the simulations with $\beta_{\rm ini}$ = 100 and $\beta_{\rm ini}$ = 400 in terms
of the distribution of $\delta_p$. In the former, the distribution of the pressure anisotropy is approximately symmetric
around 0, whereas in the latter the pressure anisotropy reaches slightly larger positive absolute values than negative
ones. \citet{ZuHone2011a} showed that sloshing motions stretch and amplify magnetic fields regardless of the initial
field strength, but that simulations with a stronger initial magnetic field reach saturation in the field strength more
quickly than those with a weaker initial field \citep[see Section 3.2 and Figures 16 and 17 of][]{ZuHone2011a}. This
effect is also present in our simulations here, and the result is that in the $\beta_{\rm ini} = 400$ simulations, there
are more regions where $dB/dt > 0$ and hence $\delta_p > 0$ (Equation \ref{eqn:pressure_anisotropy}) than in the
$\beta_{\rm ini} = 100$ simulations ($\sim$61\% vs. 55\% of the volume, respectively).
    
The other panels in Figure \ref{fig:beta_phase} show the effect of the limiters. The middle panels,
showing the simulations with ``Limiters v1,'' show that the mirror and oblique firehose instabilities place significant
restrictions on the pressure anisotropy for $\beta \gtrsim 100-1000$, but do not significantly change the distribution
of $\delta_p$ for lower values of $\beta$ compared to the ``Unlimited'' cases. The rightmost panels, showing the
simulations with ``Limiters v2,'' show that the ion-cyclotron instability places a very weak restriction on the pressure
anisotropy compared to the ``Unlimited'' simulations, while the parallel firehose instability has an effect that is
very similar to the oblique version. In the ``Limiters v1'' simulations, the fraction of the volume with positive
pressure anisotropy is reduced to $\sim$52/57\% in the $\beta_{\rm ini} = 100/400$ simulations, due to the effect of the
restrictive mirror hard-wall limiter. In the ``Limiters v2'' simulations, the fraction of the volume with positive
pressure anisotropy is increased to $\sim$58/65\% in the $\beta_{\rm ini} = 100/400$ simulations, due to the fact that
the ion-cyclotron instability is only mildly restrictive on the positive values of $\delta_p$ in these simulations, compared to the restriction from the parallel firehose instability on the negative values.
     
Another interesting quantity is the product $\delta_p\beta$, the ratio of the pressure anisotropy to the magnetic
pressure. Figure \ref{fig:delta_beta_hist} shows volume-weighted histograms of this quantity for all simulations at $t =
4.0$~Gyr within a range of $\delta_p\beta \in [-4, 4]$. For the ``Unlimited'' simulations, the distribution of
$\delta_p\beta$ peaks at $\sim 0$, with tails extending to both large positive and large negative values, with a slight
excess of positive values, as already noted in the description of Figure \ref{fig:beta_phase}. 

In the simulations with limiters, we see that the effect of the mirror and two firehose limiters is not only to restrict
$\delta_p\beta$ to the range defined by their thresholds (shown by the dashed vertical lines in Figure
\ref{fig:delta_beta_hist}), but also to shift the distribution of $\delta_p\beta$, so that there are peaks in the
distribution of $\delta_p\beta$ near these limits. For the case of the mirror instability, a non-negligible fraction of
the volume can be at its threshold of $\delta_p\beta \approx 1$ (shown by the dashed vertical lines in Figure
\ref{fig:delta_beta_hist}). In the $\beta_{\rm ini} = 100/400$ simulations, $\sim$8/13\% of the volume is at the mirror
threshold. This is not only significant for estimating the effective viscosity in the ICM, but also potentially
significant for cosmic-ray transport in the ICM, since the mirror instability is expected to be a significant source of
pitch-angle scattering for cosmic rays \citep{Ewart2024,Reichherzer2025}.

\subsubsection{Spatial Distribution of the Pressure Anisotropy}\label{sec:paniso_space}

Next, we examine the spatial distribution of the pressure anisotropy in the simulated cluster core. Slices through the
cluster center of the pressure anisotropy $\delta_p$ at $t = 3.0$~Gyr are shown in the first and third rows of panels of
Figure \ref{fig:delta_ratio} for both values of $\beta_{\rm ini}$. We see in the $\beta_{\rm ini} = 100$ simulations
(first row of Figure \ref{fig:delta_ratio}) that the pressure anisotropy has a random pattern that is driven entirely by
the turbulent velocity and magnetic fields in the cluster core. Near the CFs, where the magnetic field is increasing due
to shear amplification, the pressure anisotropy is usually positive (red). We also see regions of negative pressure
anisotropy where the magnetic field strength is decreasing due to adiabatic expansion of the CF regions (blue). There is
a region near the cluster core where the absolute value of $\delta_p$ is near zero, where the gas motions are very low
because the turbulent motions have dissipated and the entropy of the core region has flattened due to gas mixing
\citep{ZuHone2010,ZuHone2011b}, so no buoyancy forces are operating to drive further motions. This region of nearly zero
pressure anisotropy becomes larger as the size of the flat entropy region increases with time. This effect is a
consequence of the non-radiative nature of our simulations---in real clusters, gas cooling will prevent the formation of
a flat entropy core as the gas cools and sinks to the center of the cluster until it is reheated by the central AGN. The
distribution of the pressure anisotropy in the $\beta_{\rm ini} = 100$ simulations is relatively insensitive to which
limiters are applied---the stronger magnetic field overall in these simulations ensures that for most of the volume the
pressure anisotropy never reaches them (consistent with what is shown in Figure \ref{fig:beta_phase}). The $\beta_{\rm
ini} = 400$ simulations (third row of Figure \ref{fig:delta_ratio}) are similar, but the effect of the initially weaker
magnetic field is clear. As previously noted, in these simulations, the motions produce greater $dB/dt$, which results in
a larger fraction of the volume with positive pressure anisotropy. Also, the effect of the limiters is more noticeable
in these simulations, as the pressure anisotropy is more likely to reach the limits imposed by the plasma instabilities.
The effect of the limiters is most noticeable outside the central region, where the magnetic field has not yet been
amplified significantly.
        
The second and fourth rows of panels in Figure \ref{fig:delta_ratio} show a quantity known as the ``anisotropy
parameter'' \citep{Squire2023}:
\begin{equation}
\Theta \equiv 1 + \frac{4\pi\Delta{p}}{B^2},
\label{eqn:theta}
\end{equation}
This quantity is the factor by which, due
to the effect of the pressure anisotropy, the ``effective magnetic tension'' in the ICM is larger or smaller than the true tension. The left-most panels of the second and fourth rows of Figure
\ref{fig:delta_ratio} show slices of $\Theta$ without limiters applied to $\Delta{p}$. A significant portion of the
volume has positive $\Theta$ with a value close to unity (light green), indicating that the corrective effect of the
pressure anisotropy is mild. Nevertheless, there are some regions with either very large positive values of $\Theta$
(dark green) or very large negative values (brown). The absolute value of $\Theta$ in these regions can get very large;
up to $\sim 10^2-10^3$, indicating either a very large positive or negative effective magnetic tension. These extreme
values appear where the magnetic field is changing most rapidly, viz., near the CF surfaces.

In reality, such large values of $\Theta$ should not occur because of plasma instabilities. If limiters on the pressure
anisotropy due to these instabilities are applied, the distribution of $\Theta$ changes significantly. The middle panels
of the second and fourth rows of Figure \ref{fig:delta_ratio} show the effect of the ``v1'' limiters, enforcing the oblique-firehose and mirror thresholds. By
construction, the distribution of $\Theta$ is more restricted, with the vast majority of the volume having $\Theta$
close to unity, indicating that the viscous stress never gets significantly larger than the stress from the magnetic
tension. In the case where the ``v2'' limiters are applied, $\Theta$ can reach extremely large positive values of $\sim
10^2-10^3$ because the ion-cyclotron instability is far less restrictive in this regime than the mirror instability.
However, in this case $\Theta$ is bounded at the lower end by zero due to the parallel firehose instability, at whose threshold the magnetic tension $B^2/4\pi$ and the pressure anisotropy $\Delta{p}$ exactly cancel (white
regions), predominantly outside the sloshing region.

Figure \ref{fig:limiter_maps} provides a closer look at the regions of the cluster core where the pressure anisotropy is close to the hard-wall limits in the ``Limiters, v1'' simulations. The red regions indicate where the pressure anisotropy is at the mirror limit and the blue regions indicate where it is at the oblique-firehose one. In general, the regions at either limit occupy a small fraction of the volume, distributed mostly outside the sloshing region and in between parts of the sloshing region where the magnetic field has not been amplified as strongly. Consistent with the previous discussion, the $\beta_{\rm ini} = 400$ simulation has a larger fraction of the volume at either limit, and there are more regions at the mirror limit than at the oblique-firehose one.

\begin{figure}[!t]
\centering
\includegraphics[width=0.49\textwidth]{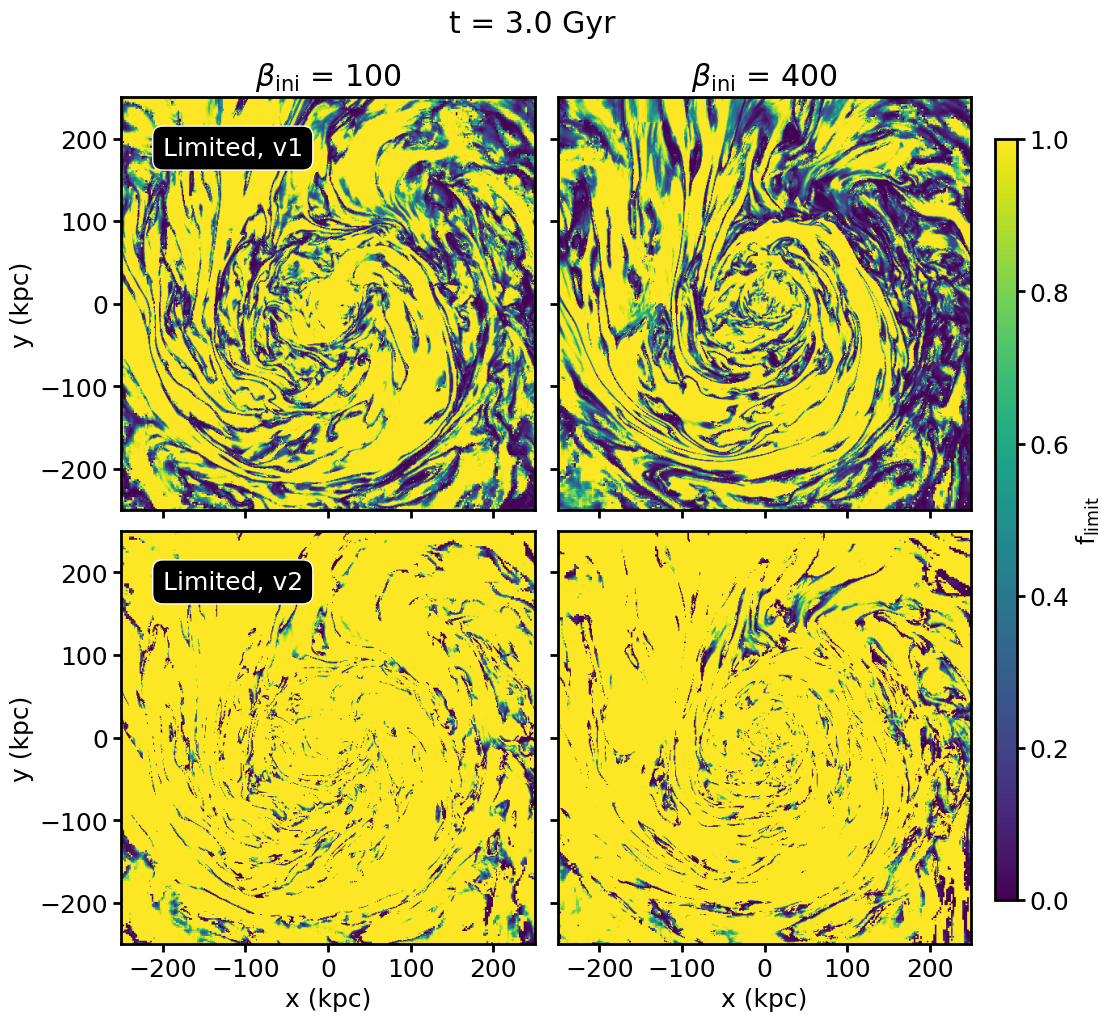}
\caption{Suppression of viscous momentum flux by plasma instabilities. Panels show slices through $f_{\rm limit}$, defined in Equation \ref{eqn:flimit}, for the simulations where plasma-instability-based limiters are applied at $t$ = 3.0~Gyr.\label{fig:suppression}}
\end{figure}

\subsection{Effective Limits on the Viscous Flux}\label{sec:visc_limits}

As noted in Section \ref{sec:intro}, the viscosity in the ICM has often been estimated indirectly from the properties of
features seen in X-ray SB, such as CFs and SB fluctuations. Such estimates are limited to the simplified
model of a (potentially) suppressed isotropic viscosity, given that direct constraints on the magnetic-field lines'
direction and structure in a particular location of a cluster's atmosphere are difficult to achieve. It is therefore
instructive to determine what the effective viscous momentum flux is from our simulations under such a simplified assumption.  

First, the geometry of the magnetic field itself places limits on the viscous flux simply due to the anisotropic nature
of Braginskii viscosity. A straightforward way to quantify this is to consider Equation \ref{eqn:pressure_anisotropy2}
and determine for what value of $\bhat$ the pressure anisotropy is maximized. Not coincidentally, this is achieved by
finding the eigenvector $\bhat_{\rm max}$ corresponding to the largest eigenvalue of the following tensor:
\begin{equation}
\frac{1}{2}[\nabla{\bf v} + (\nabla{\bf v})^\mathsf{T}] - \frac{1}{3}(\nabla\cdot{\bf v})\identity,
\end{equation}
which is proportional to the viscous-stress tensor (Equation \ref{eqn:visc_tensor_hydro}) in the isotropic case. We can then compute the viscous-flux suppression factor relative to the isotropic case as
\begin{equation}
f_{\rm geom} = \frac{(\bhat\bhat - \identity/3):\nabla{\bf v}}{(\bhat_{\rm max}\bhat_{\rm max} - \identity/3):\nabla{\bf v}}.
\label{eqn:fgeom}
\end{equation}
In the incompressible limit, the numerator in Equation \ref{eqn:fgeom} simply reduces to $\bhat\bhat:\nabla{\bf v}$,
which has been used by previous authors \citep{StOnge2020,Majeski2024} to demonstrate the dynamical suppression of
viscosity in high-$\beta$, collisionless plasma turbulence. As a reference point, in the case of an isotropically
distributed random magnetic field, the angle-averaged suppression of viscosity is $f_{\rm geom} = 1/5$
\citep{Malyshkin2002,Nulsen2013}.

\begin{figure}[!t]
\centering
\includegraphics[width=0.49\textwidth]{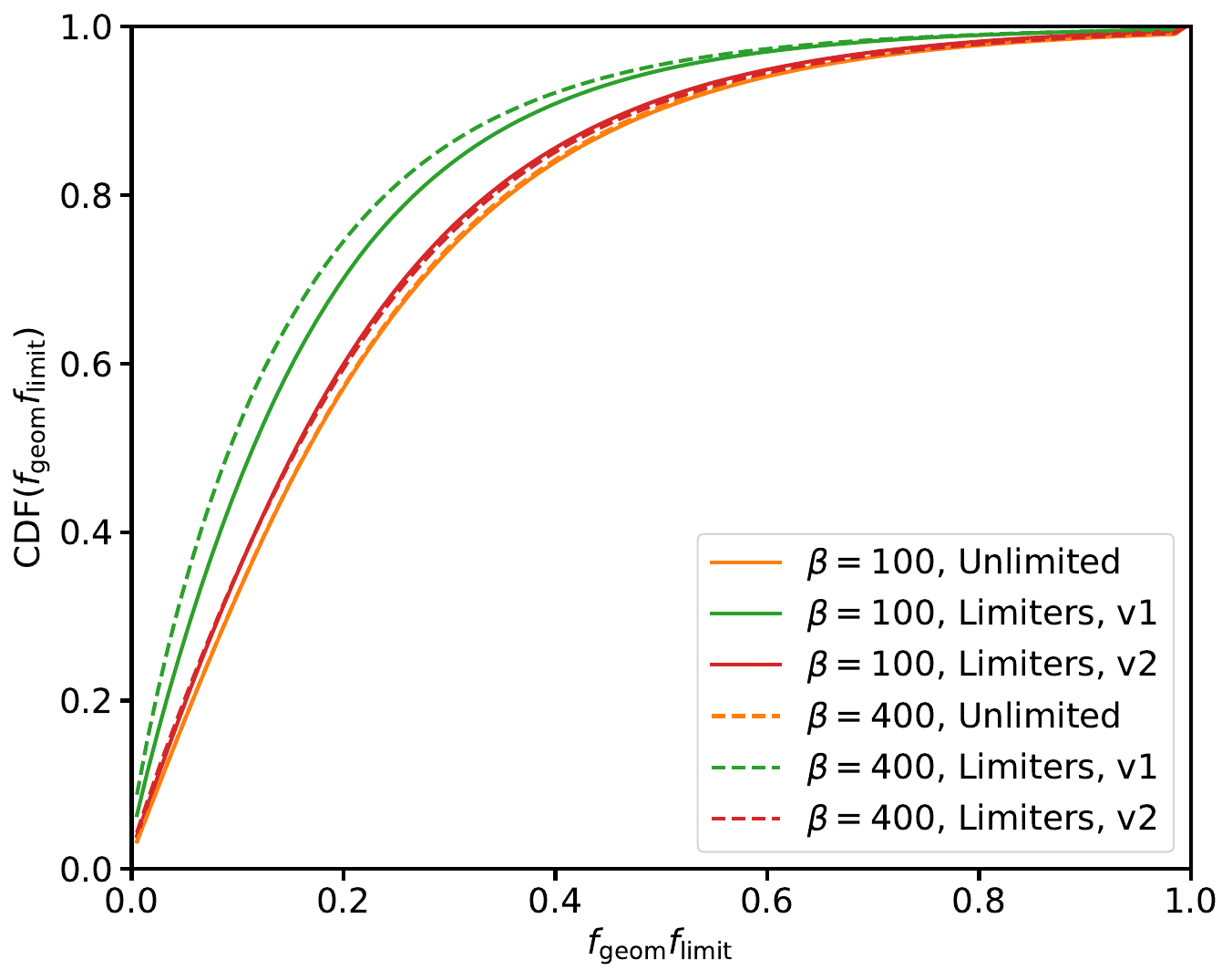}
\caption{Cumulative histogram of the quantity $f_{\rm limit}f_{\rm geom}$ within a radius of 150~kpc from the cluster center, calculated from all the simulations with viscosity, at $t$ = 3.0~Gyr.\label{fig:supp_hist}}
\end{figure}

Figure \ref{fig:fgeom} shows slices of $f_{\rm geom}$ through the center of the simulation domain for the $\beta_{\rm
ini} = 100$ and $\beta_{\rm ini} = 400$ simulations at $t$ = 3.0~Gyr (the result is very similar for other epochs). From
these, we can see that the effect of the anisotropy on reducing the viscous flux can be substantial, whether the limits
imposed by plasma instabilities are included in the simulation or not. The effect is most pronounced in the regions
where the magnetic field is most tangled, whereas in regions where velocity gradients align with the magnetic field
direction (e.g., where velocity shears are stretching and amplifying the fields) the suppression is minimal. The degree
of geometric viscosity suppression appears in many regions to exceed the angle-averaged value of $1/5$, which may be a
signature of the self-organization process known as magneto-immutability~\citep{squire2019,Squire2023,Majeski2024}. In
Section \ref{sec:immutability}, we will discuss the flow-field signatures and consequences of this process in more detail.

\begin{figure*}
\centering
\includegraphics[width=0.98\textwidth]{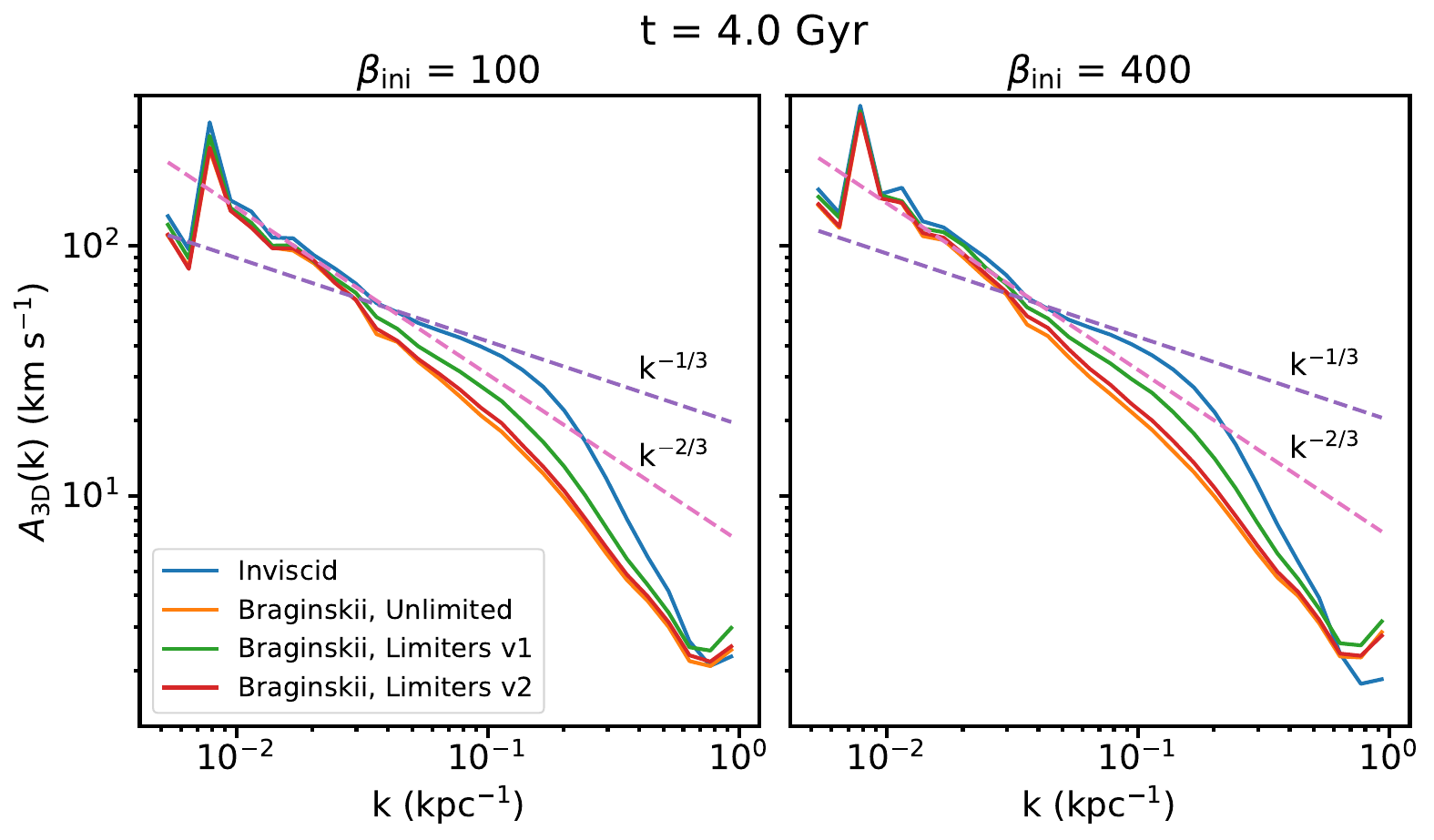}
\caption{Velocity-amplitude spectra $A_{\rm 3D}(k)$ for all of the simulations in the sample at $t$ = 4.0~Gyr, shown with solid curves. Dashed lines for power-law slopes of -1/3 (Kolmogorov) and -2/3 are shown for comparison.\label{fig:vas}}
\end{figure*}

The second form of suppression of viscous flux comes from the plasma microinstabilities. To quantify this, we
simply note that the absolute value of the pressure anisotropy is limited by these instabilities compared to its maximum value by a factor of
\begin{equation}
f_{\rm limit} = \frac{|\Delta{p}|}{3\mu|{S}|},
\label{eqn:flimit}
\end{equation}
where $S$ is the parallel rate of strain, previously defined in Equation \ref{eqn:pressure_anisotropy2}. Figure
\ref{fig:suppression} shows slices through the center of the domain of $f_{\rm limit}$ for all of the simulations where
limiters on the pressure anisotropy are applied at $t = 3.0$~Gyr (the results are similar for other epochs). For the
``Limiters, v1'' simulations (top panels), the viscous flux is hardly suppressed at all inside the region bounded by the
CFs because here the magnetic field is strongest, and it is more difficult for the pressure anisotropy to reach the
limits. Outside this region, the viscosity can be significantly suppressed, depending on the initial magnetic-field
strength ($\beta_{\rm ini}$ = 100 vs. 400). For the ``Limiters, v2'' simulations (bottom panels), since most of the
inner region has a positive pressure anisotropy and the ion-cyclotron instability is not as stringent as the mirror
instability, the suppression of the viscous flux is not as significant as in the ``Limiters, v1'' simulations.

Combining these two factors as $f_{\rm limit}f_{\rm geom}$, we can estimate the effective ``isotropic'' viscosity in the
cluster core region. Figure \ref{fig:supp_hist} shows cumulative histograms of $f_{\rm limit}f_{\rm geom}$ for a
spherical volume with radius 150~kpc centered on the cluster center, for all of our simulations with viscosity, at $t$ =
3.0~Gyr. From these curves, we can observe that a significant fraction of this inner volume has a suppressed effective
viscosity: $\sim$57-75\% of the volume has $f_{\rm limit}f_{\rm geom}$ less than 0.2, and $\sim$90-96\% of the volume
has $f_{\rm limit}f_{\rm geom}$ less than 0.5.

\subsection{Properties of the Velocity Field}\label{sec:velocity_results}

\subsubsection{Velocity Amplitude Spectra}\label{sec:turb_spectra}

\begin{figure*}
\centering
\includegraphics[width=0.48\textwidth]{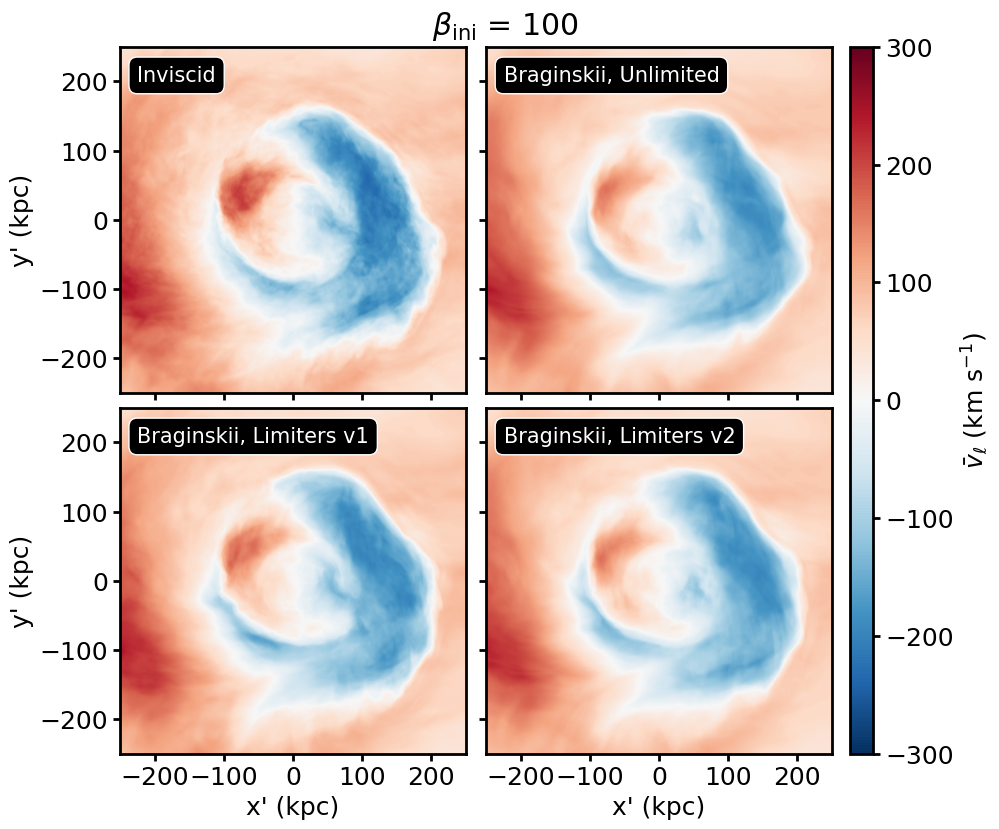}
\includegraphics[width=0.47\textwidth]{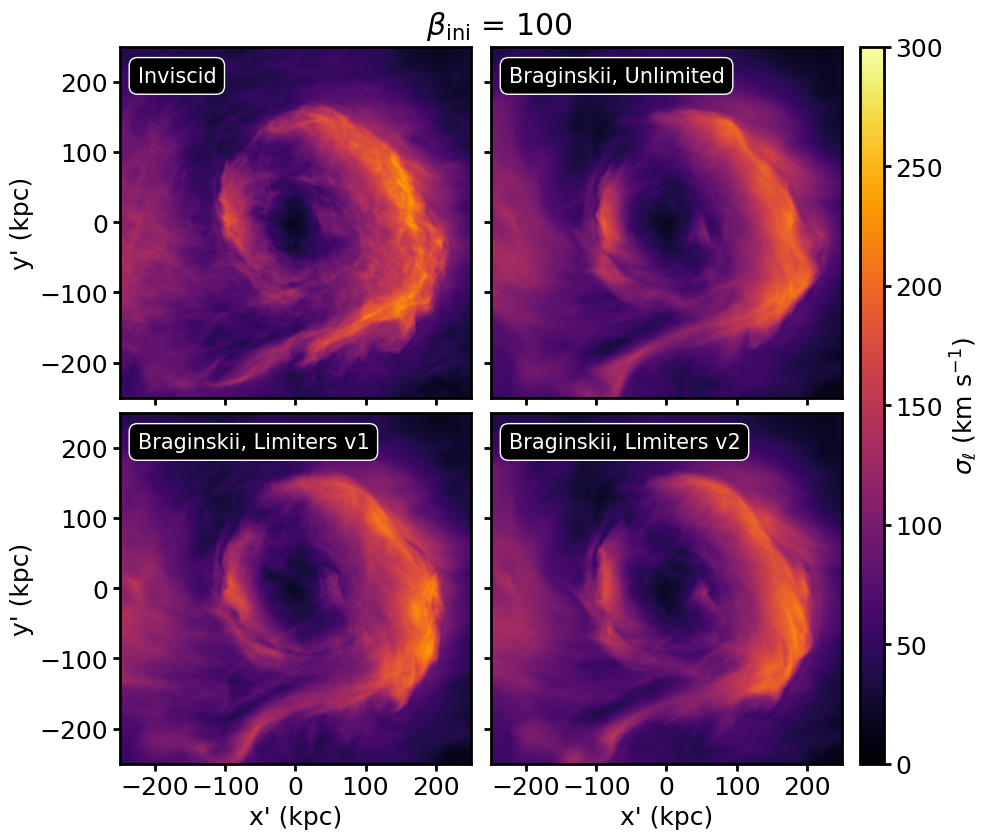}
\includegraphics[width=0.48\textwidth]{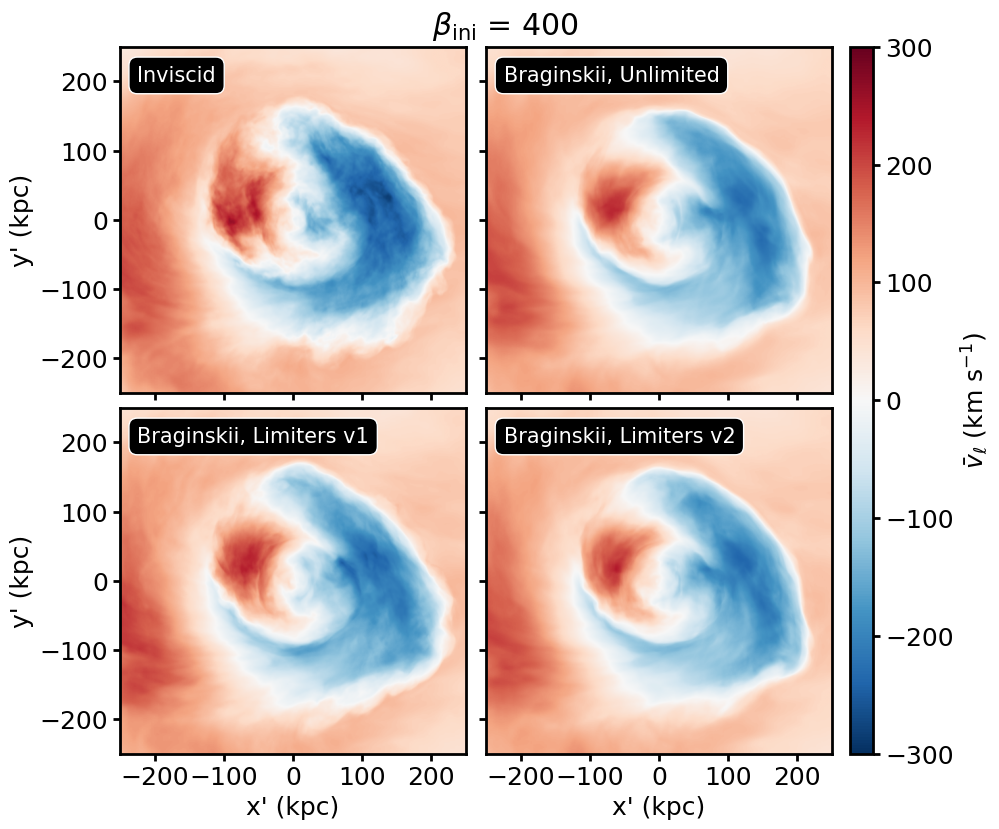}
\includegraphics[width=0.47\textwidth]{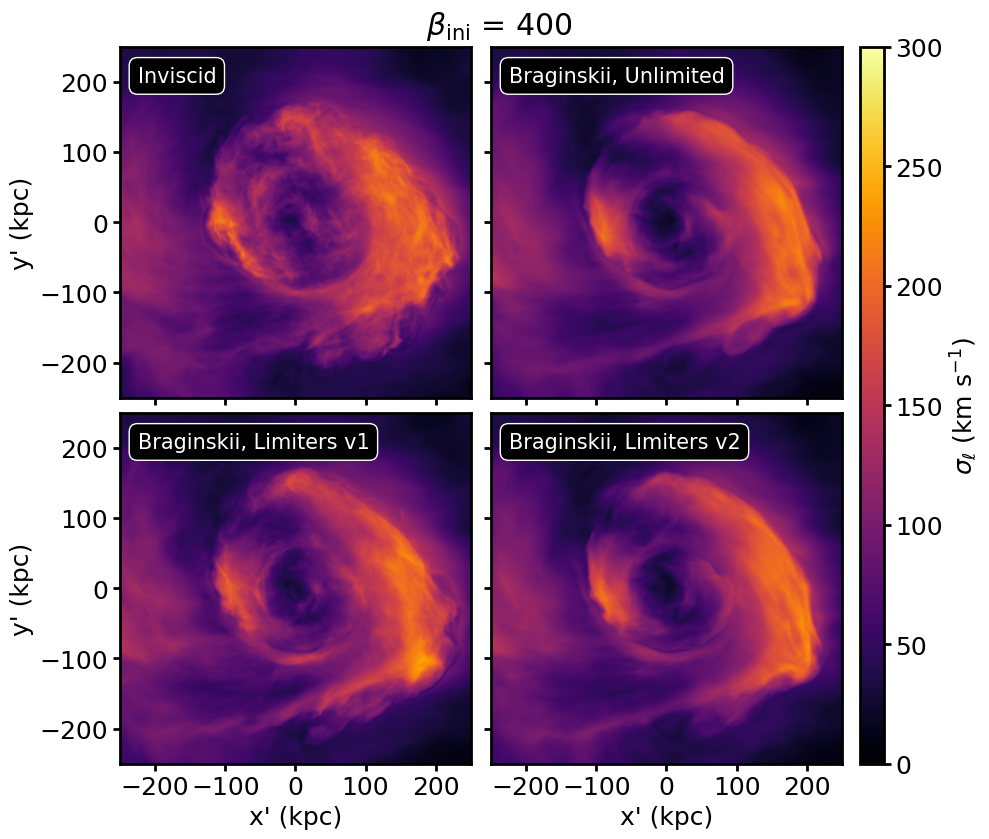}
\caption{Maps of the projected bulk velocity ${\bar v}_\ell$ (left panels) and velocity dispersion $\sigma_\ell$ (right
panels) for all of the simulations ($\beta_{\rm ini}$ = 100 on top and $\beta_{\rm ini}$ = 400 on the
bottom). The line-of-sight direction $\hat{\boldsymbol\ell}$ for the projection is in the $x-z$ plane of the simulation
domain, 45$^\circ$ between the $x$ and $z$-axes.\label{fig:proj_maps}}
\end{figure*}

We construct velocity-amplitude spectra from our simulations to study the effect of the different prescriptions for
viscosity on different length scales. The velocity-amplitude spectrum can be derived from the velocity power spectrum,
and gives the amplitude $A_{\rm 3D}(k)$ of the velocity at a given length scale $1/k$ for a given wavenumber
$k$.\footnote{Note that, here and throughout this work, for the purposes of visualization, we define the wavenumber $k$
as the inverse of the corresponding length scale $\ell$, without the customary factor of 2$\pi$, to aid the reader in
relating the wavenumber to the physical length scale.} To compute $A_{\rm 3D}$, we use the velocity field ${\bf v}({\bf
x})$ within the volume $V$ = (250~kpc)$^{3}$ centered on the cluster center and take its 3D Fourier transform to obtain
$\tilde{\bf v}({\bf k})$, where ${\bf k}$ is the wavevector that gives the 3D power spectrum $P_{\rm 3D}({\bf k}) =
|\tilde{\bf v}({\bf k})|^2$. Before doing this, we multiply the input ${\bf v}({\bf x})$ by a Tukey window function to
suppress the effects of the non-periodic boundaries of the extraction region on the power spectrum. By averaging $P_{\rm
3D}({\bf k})$ over spherical shells in $k$-space, we obtain $P_{\rm 3D}(k)$, from which we can define $A_{\rm 3D}(k)$ to
be \citep{Zhuravleva2012}: 
\begin{equation}
A_{\rm 3D}(k) = \sqrt{4{\pi}{k^3}P_{\rm 3D}(k)}.
\end{equation}

\begin{figure*}
\centering
\includegraphics[width=0.98\textwidth]{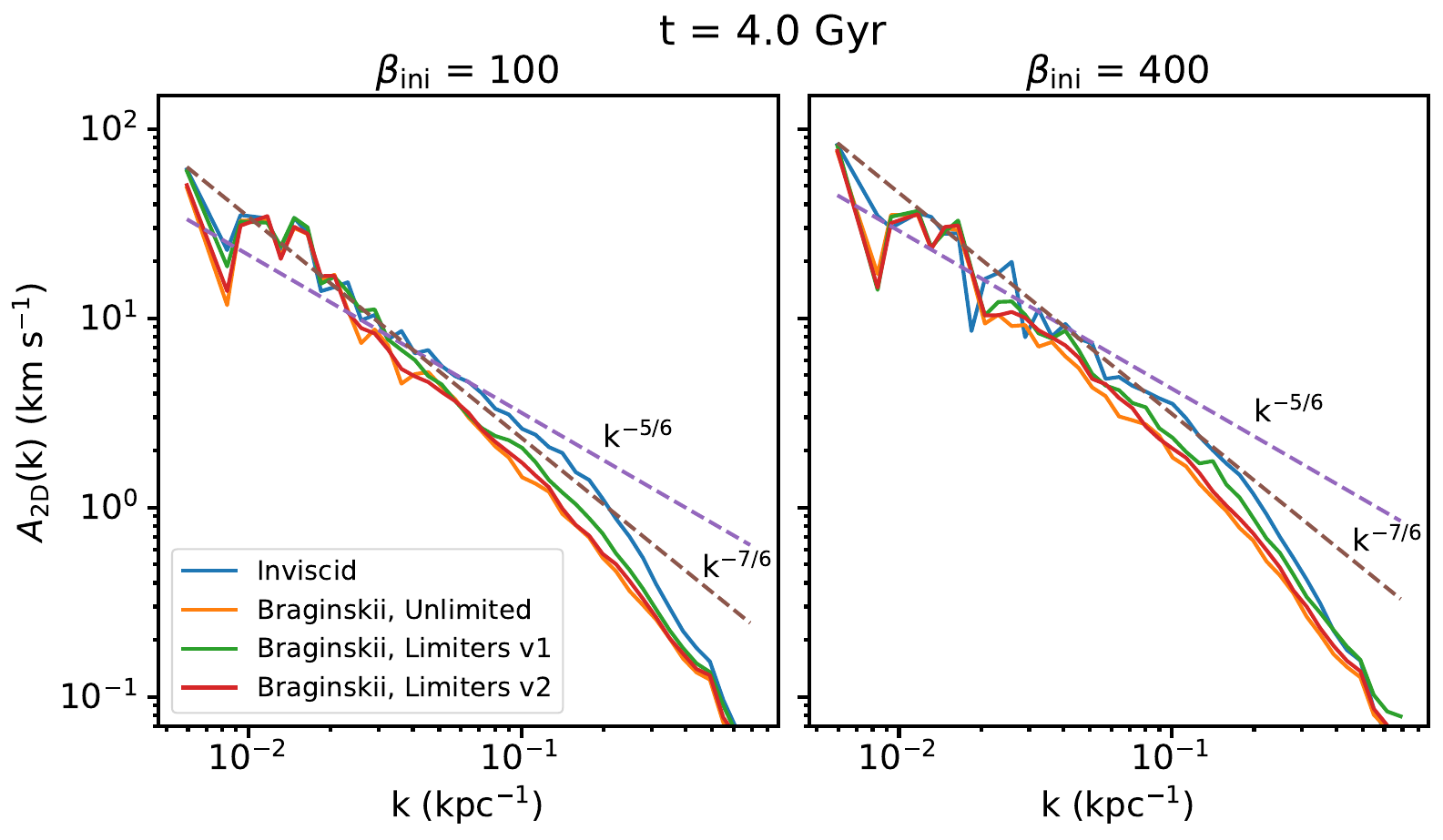}
\caption{Projected bulk-velocity-amplitude spectra $A_{\rm 2D}(k)$ for all of the simulations in the sample, at $t$ = 4.0~Gyr. Dashed lines for power-law slopes of -5/6 (Kolmogorov) and -7/6 are shown for comparison.\label{fig:vas_2d}}
\end{figure*}

Figure \ref{fig:vas} shows the velocity-amplitude spectra $A_{\rm 3D}(k)$ obtained using this method for all of the
simulations in the sample at the epoch $t = 4.0$~Gyr. At wavenumbers $k \lesssim$ 0.03~kpc$^{-1}$, $A_{\rm 3D}(k)$ has a
steep slope in all simulations, close to $-2/3$, which is steeper than the Kolmogorov slope of $-1/3$. For wavenumbers
between 0.03~kpc$^{-1} \lesssim k \lesssim$ 0.2~kpc$^{-1}$ the slope varies between the simulations, depending on
whether or not viscosity is included. In the inviscid simulations (blue curves) this range has a nearly power-law shape
with slope close to $-1/3$, consistent with the Kolmogorov spectrum of turbulence. For the simulations with unlimited
Braginskii viscosity (orange curves), the slope of the spectrum is steeper, closer to $\sim$~-2/3 to -1. At $k \sim$
0.2~kpc$^{-1}$, the ``Unlimited'' simulations with the strongest viscosity have a velocity amplitude a factor $\sim$2
smaller than in the inviscid case. The ``Limiters, v1'' simulations (green curves) are between these two cases, but the
``Limiters, v2'' simulations are very close to the ``Unlimited'' simulations, given that the positive pressure
anisotropy is not strongly suppressed in this case. At wavenumbers $k \gtrsim$ 0.2~kpc$^{-1}$, corresponding to a length
scale of 5~kpc, or 5$\Delta{x}$, the slope of the spectrum steepens significantly in all of the simulations, which is a
signature of the dissipation of gas motions by numerical viscosity. The characteristics of the velocity amplitude
spectra in these simulations do not appear to have a noticeable dependence on the initial plasma $\beta$. 

The steeper slope in all simulations at the larger length scales ($\gtrsim$ 50~kpc) is plausibly explained by the bulk
sloshing motions. These are large-scale coherent flows (which can decay to turbulence) that arise from buoyancy
oscillations of parcels of the ICM at the local Brunt-V\"ais\"al\"a (BV) frequency, which varies with radius in the
cluster atmosphere due to the gravitational potential and entropy stratification. The varying BV frequencies lead to an
outwards-moving coherent wave pattern, as first roughly described by \citet{Churazov2003}, demonstrated in simulations
by \cite{AM06}, and recently explored in detail by \citet{Roediger2024} \citep[see also][]{Choudhury2025}. These
motions have been shown in previous works \citep{Vazza2012,ZuHone2013,ZuHone2015} to have a steep velocity spectrum, if they have not had time to decay into turbulence.

In the intermediate range of length scales (5-50~kpc), though it is obvious that viscosity is responsible for the
steeper slopes observed in the velocity amplitude spectra, the physical reason for the particular value of the slope
($\sim$~-2/3 to -1) is not clear. Given that the slope of the part of the spectrum dominated by sloshing has a similar
value, it is possible that viscosity suppresses the development of KHI and subsequent turbulence, and thus the
velocity-amplitude-spectrum at all scales becomes dominated by the sloshing motions. Alternatively, similarly steep
slopes have been seen in previous works which explore driven MHD turbulent dynamos in homogeneous domains, in particular
in the saturated regime \citep[see, e.g.][]{Schekochihin2004,StOnge2020,Grete2021}, though so far no satisfying
explanation for this behavior has been offered. 

\begin{figure*}
\centering
\includegraphics[width=0.98\textwidth]{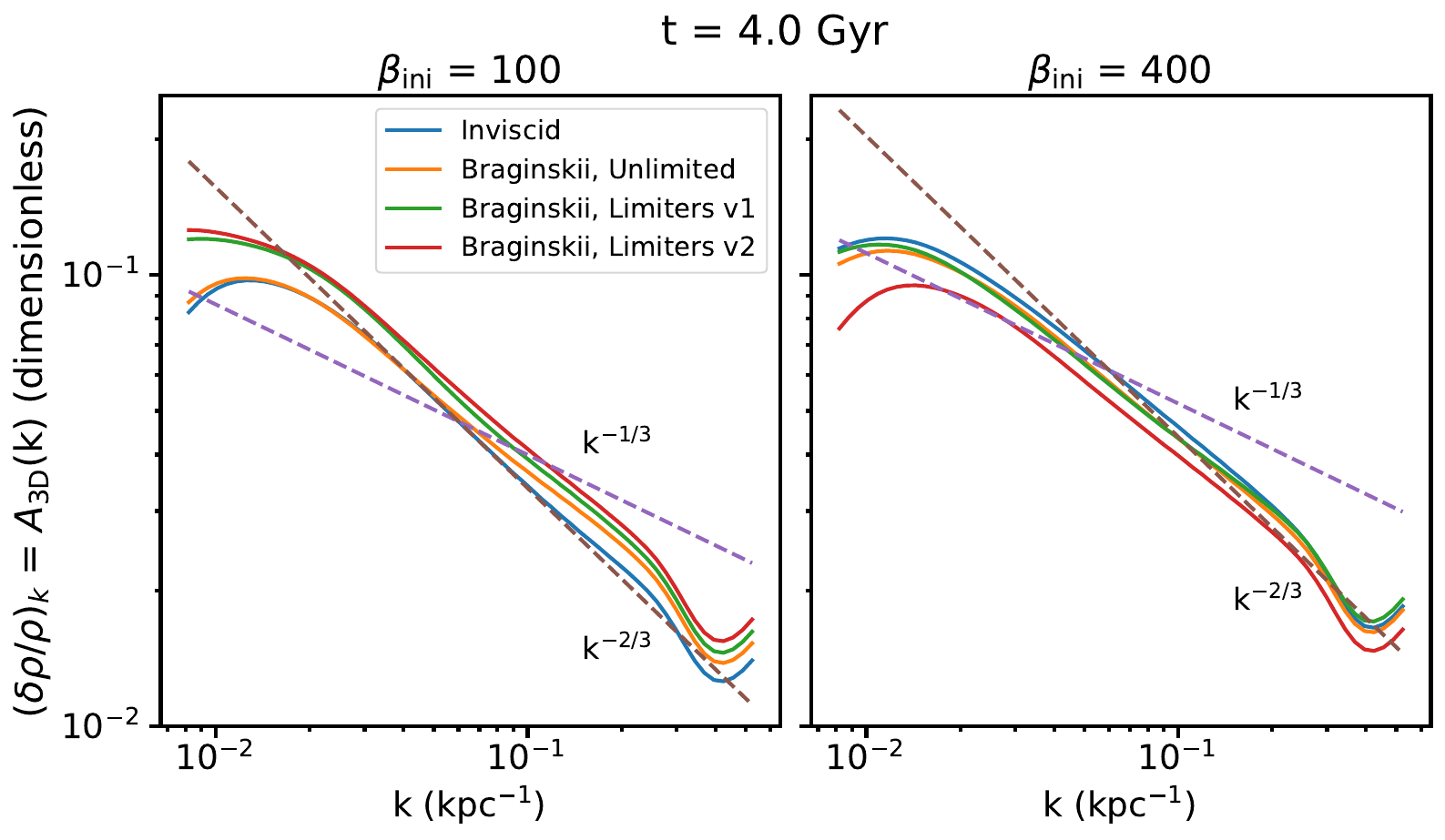}
\caption{3D density-fluctuation-amplitude spectra $A_{\rm 3D}(k)$ for all of the simulations in the sample, at $t$ = 4.0~Gyr. Dashed lines for power-law slopes of -1/3 (Kolmogorov) and -2/3 are shown for comparison.\label{fig:dsb_spectra}}
\end{figure*}

\subsubsection{Projected Velocity Fields}\label{sec:proj_vel}

Like all other quantities obtained from astronomical observations, the velocity field of the ICM is measured in a 2D
projection of the full 3D field. With microcalorimeter instruments like {\it XRISM}'s Resolve, velocities can be
measured from the Doppler shifting and broadening of emission lines, giving the average (or ``bulk'') velocity ${\bar
v}$ and the velocity dispersion $\sigma$. To produce maps of these projected quantities from our simulations, we
integrate the component $v_{\ell}$ of the 3D velocity field along the sight line element $\ell$,
weighted by the normalized X-ray emissivity $\tilde{\varepsilon}$ in the 6.0-8.0~keV band (the energy range containing
the prominent Fe emission lines used to measure Doppler shifting and broadening by {\it XRISM}):
\begin{eqnarray}
{\bar v}_{\ell}({\bf x}') &=& \displaystyle\int{\tilde{\varepsilon}({\bf x})}v_{\ell}({\bf x})d{\ell}, \\
\sigma_{\ell}^2({\bf x}') &=& \displaystyle\int{\tilde{\varepsilon}({\bf x})}v_{\ell}^2({\bf x})d{\ell}-{\bar v}_{\ell}^2({\bf x}'),
\end{eqnarray}
where ${\bf x}$ is the 3D position within the cluster and ${\bf x}'$ is the 2D projected sky position. In what follows,
we choose a line-of-sight direction $\hat{\boldsymbol\ell}$ in the $x-z$ plane of the simulation box, at an angle of
45$^\circ$ to both the $x$ and $z$-axes, which ensures that a significant portion of the sloshing motions are projected
onto the observer's sight line.

Maps of ${\bar v}_{\ell}$ and $\sigma_{\ell}$ for all of the simulations are presented in Figure \ref{fig:proj_maps}. On
large scales, the distribution of bulk velocities (line shifts, left panels) and velocity dispersions (line widths,
right panels) is very similar for simulations with the same $\beta_{\rm ini}$. This is consistent with previous results
using idealized simulations of gas sloshing by \citet{ZuHone2016,ZuHone2018}, which showed that even in an unrealistic
scenario with isotropic Spitzer viscosity, the large-scale velocity distributions are very similar to the inviscid case.
What is noticeable is that the smaller-scale variations in both ${\bar v}_{\ell}$ and $\sigma_{\ell}$ are smoothed out
in simulations with higher effective viscosity. As expected from the 3D power spectra shown in Figure \ref{fig:vas}, the
simulations with the smoothest velocity fields are the ``Braginskii, Unlimited'' cases, whereas including limiters on
the pressure anisotropy (and hence on the viscous stress) results in more structure at smaller scales. 

Following on this, we can construct velocity-amplitude spectra for the 2D line-shift map in a similar way as we did for
the 3D velocity distribution. In this case, we construct the power spectrum $P_{\rm 2D}$ of the line shift ${\bar
v}_\ell$ from the maps shown in Figure \ref{fig:proj_maps}, and compute the amplitude spectrum as
\begin{equation}
A_{\rm 2D}(k) = \sqrt{2{\pi}{k^2}P_{\rm 2D}(k)},
\end{equation}
where $k$ is now the magnitude of the 2D wavenumber. We use an area of (250~kpc)$^2$, corresponding to the same
width as was used for the 3D spectra in the previous section. The results are shown in Figure \ref{fig:vas_2d}. As shown
in \citet{Zhuravleva2012}, for sufficiently large values of $k$ (depending on the emission-measure profile of the
cluster), the 2D and 3D amplitude spectra are related by $A_{\rm 2D}(k) \propto A_{\rm 3D}(k)k^{-1/2}$ (their Equation
22), implying $A_{\rm 2D}(k) \propto k^{-5/6}, k^{-7/6}$ for $A_{\rm 2D}(k) \propto k^{-1/3}, k^{-2/3}$ (see Figure
\ref{fig:vas}). Figure \ref{fig:vas_2d} confirms this expectation to a certain extent, where the $A_{\rm 2D}(k)$ for the
``Inviscid'' simulations has a shallower slope closer to -5/6, while the simulations with Braginskii viscosity have a
steeper slope closer to -7/6, though visually it is slightly harder to distinguish the slopes in the 2D projection than
in 3D. We have also experimented with computing $A_{\rm 2D}(k)$ from similar bulk-velocity maps projected along other
sight lines and found the spectra to be very similar to those shown in Figure \ref{fig:vas_2d}. 

\subsubsection{Density-Fluctuation-Amplitude Spectra}\label{sec:v_from_sb}

\begin{figure*}[!t]
\centering
\includegraphics[width=0.48\textwidth]{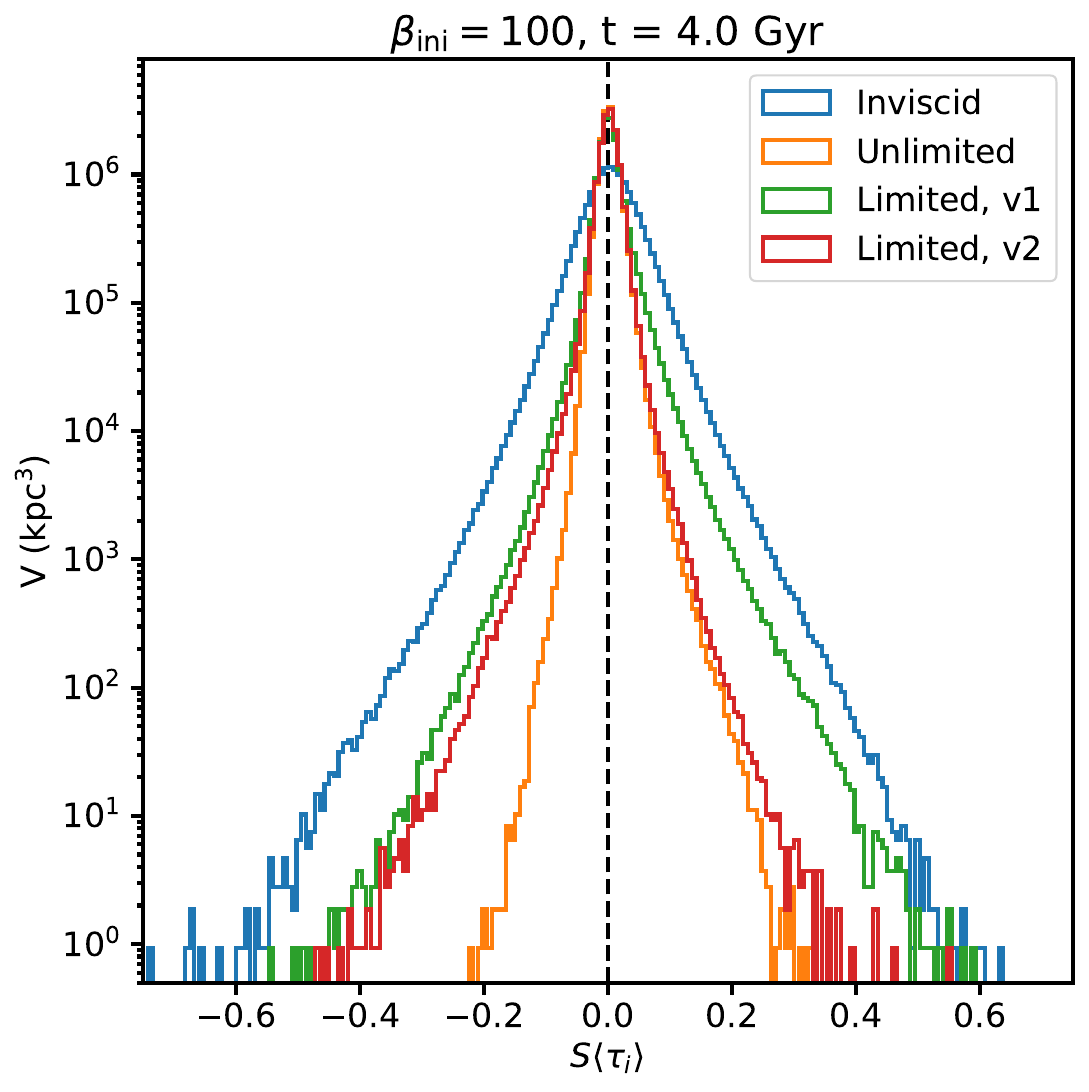}
\includegraphics[width=0.48\textwidth]{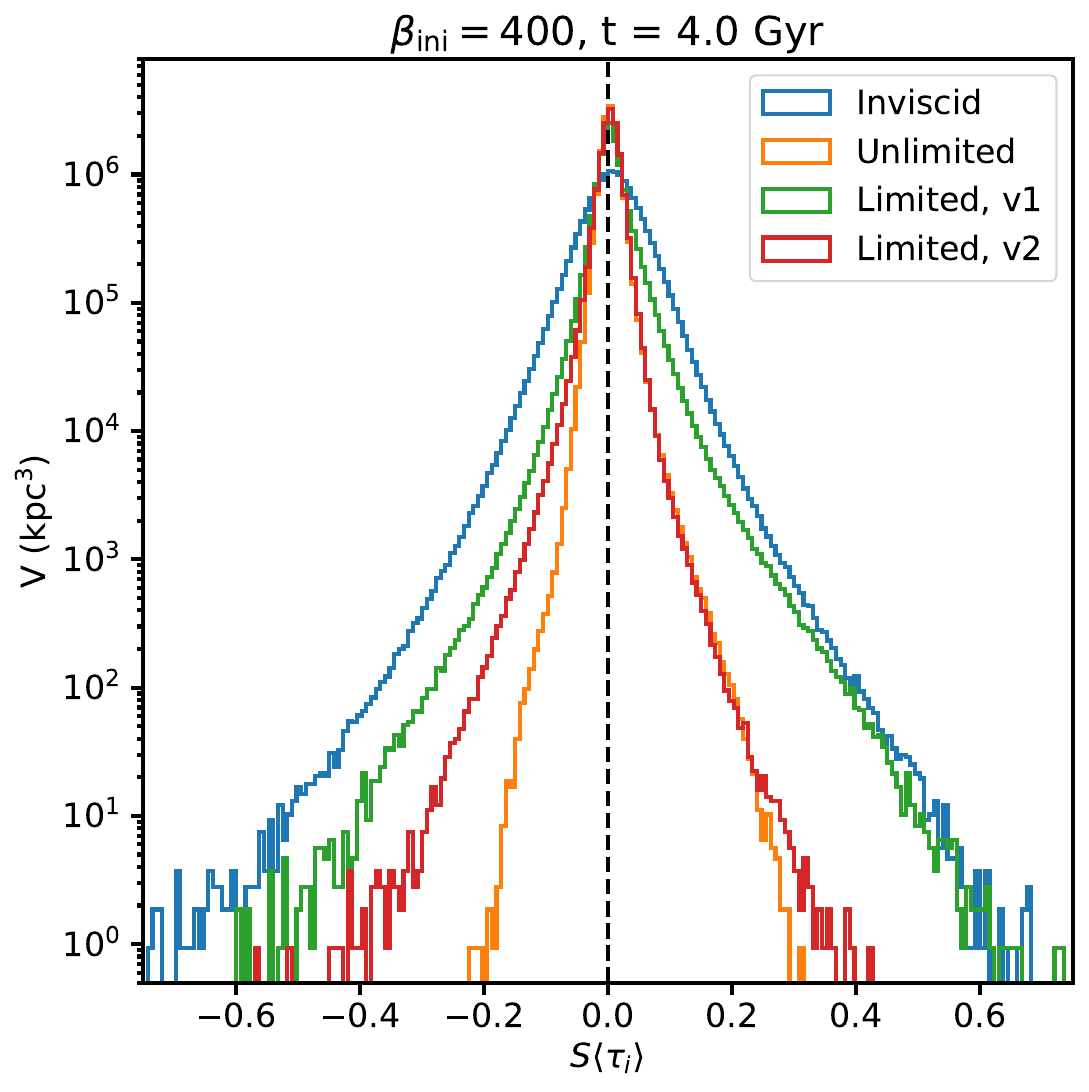}
\caption{\label{fig:immutability} Probability distribution functions (PDFs) of the field-aligned rate of strain $S =(\bhat\bhat - \identity/3):\nabla{\bf v}$ in each simulation, normalized to the ion-ion collision time $\tau_i$ averaged over the entire domain.}
\end{figure*}

The different forms of viscosity explored here may affect the ``sharpness'' of discontinuities in the spiral pattern
(Figure \ref{fig:resid_maps}). It may be possible to distinguish between these different models from the observed power
spectra of X-ray SB fluctuations. We adapt an approach used in many previous observational studies, measuring the power
spectrum of SB fluctuations via the $\Delta$-variance method \citep{arevalo2012}. For this task, we use the projected
X-ray SB residual images previously shown in Figure \ref{fig:resid_maps}. The power spectrum of SB fluctuations is
measured from these residual images in the same 250~kpc-wide region as in Sections \ref{sec:turb_spectra} and
\ref{sec:proj_vel}, then deprojected to find the 3D density-fluctuation-amplitude spectra \citep[following,
e.g.,][]{Heinrich2024,churazov2012}. The resulting spectra are shown in Figure \ref{fig:dsb_spectra}.

This method, which is tailored to observational measurements, produces a smoothed power spectrum that makes it difficult
to identify its particular features, such as the change in slope of the velocity-amplitude spectra
between different simulations, as shown in Figures \ref{fig:vas} and \ref{fig:vas_2d}. This, in combination with the
projection of fluctuations along the line of sight, makes it extremely difficult to measure a difference in the slope of
the density-fluctuation-amplitude spectrum resulting from the effects of viscosity. We experimented with projecting
along different axes and found no discernible differences. There is some indication that the density fluctuations
spectra may be more distinguishable if we focus on a smaller region to exclude the CF surfaces, but this comes at the
cost of increasing uncertainties due to the smaller sampling area.

\subsection{Flow Self-Organization}\label{sec:immutability}

%can also refer back to figure six, comparing the unlimited case with the results of MKS24, where the non-hard walls appeared to allow for better self-regulation of the pressure anisotropy, except in this case self-regulation is overall less effective that instability scattering (especially because in between the thresholds the viscous damping should naively be similarly effective for both limiters v1 and unlimited, but it isnt)

As shown in Figure \ref{fig:fgeom}, magnetic-field-induced anisotropy already limits the ability of Braginskii MHD to
produce viscous stresses. Nonetheless, in high-$\beta$ plasmas, the suppression of viscous momentum flux has, in some
cases, been found to exceed that which can be accomplished through tangled magnetic-field geometry
alone~\citep{squire2019,Squire2023}. The property of the turbulent dynamics responsible for this suppression is known as
`magneto-immutability'. Magneto-immutability results from a self-organization process in which turbulent flows are
rearranged dynamically in a manner that reduces the magnetic-field-aligned rate of strain. As a direct consequence of
this and of the induction equation \eqref{eqn:bfield}, the inertial range of magneto-immutable turbulence features
smaller fluctuations in the magnetic-field strength than is typical of MHD turbulence and a reduction of viscous
stresses -- without modifying the coefficient of kinematic viscosity itself (as the mirror, ion-cyclotron, and firehose
instabilities would do). Diagnosing this effect may be integral to developing an understanding of viscosity in the ICM
and to the interpretation of related observations because it has been predicted to extend turbulent cascades down to
microphysical scales~\citep{Majeski2024}. It is therefore interesting to evaluate the simulations presented in this work
for signatures of this self-organization process. 

\begin{figure*}
\centering
\includegraphics[width=0.95\textwidth]{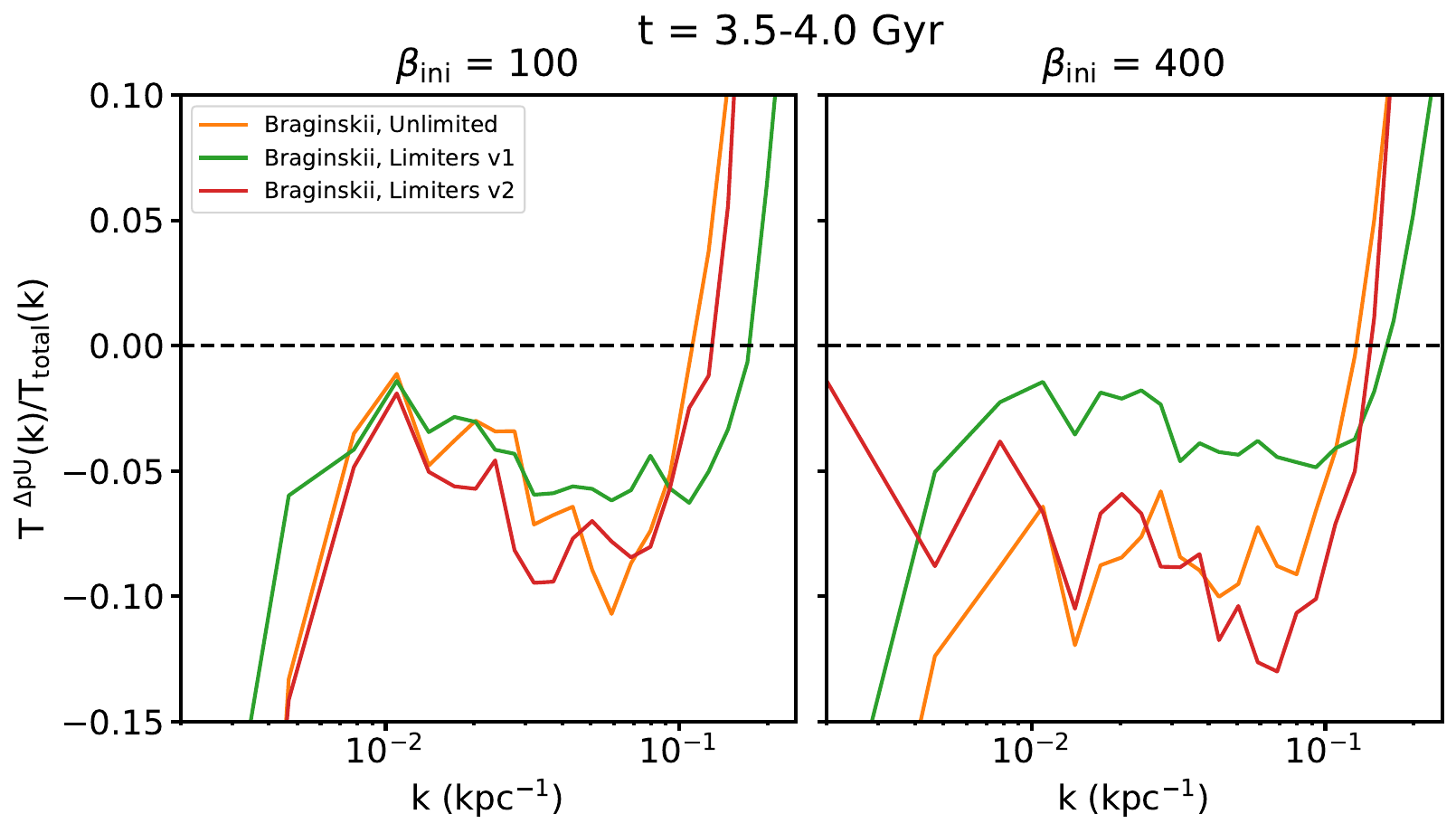}
\caption{Transfer functions ${\rm T}^{\Delta{p}\rm{U}}(k)$ from kinetic to internal energy due to pressure-anisotropy stress, plotted as functions of wavenumber $k$ and normalized by the total energy-cascade rate. They are calculated from the simulations by averaging between the epochs of 3.5-4.0~Gyr.\label{fig:transfer_function}}
\end{figure*}

The physics that enables magneto-immutability is the feedback of anisotropic pressure stresses in the momentum equation
\eqref{eqn:momentum}. For this reason, inviscid MHD models cannot be magneto-immutable: the rate of strain's components
associated with the Braginskii viscous stress, $S = (\bhat\bhat - \identity/3):\nabla{\bf v}$, can only be suppressed in
such simulations via the tangled geometry of the magnetic-field lines. A suppression of $S$ beyond that which is already
present in inviscid MHD can then be viewed as a signature of magneto-immutability. In Figure \ref{fig:immutability}, we
compare the probability distribution functions (PDFs) of $S$ for each of our simulations (normalized by the average
ion-ion collision timescale $\tau_i$), including the inviscid MHD run. It is immediately clear that the inviscid
simulation features a PDF with greater variation in the field-aligned rate of strain than any of the Braginskii
simulations, regardless of the limiter scheme used. The narrowest of these distributions belongs to the unlimited
Braginskii simulation, because the hard-wall limiters employed in other Braginskii runs reduce the strength of the
Braginskii stress and therefore induce MHD-like behavior whenever they are active. With increasingly restrictive
limiters, the distributions are therefore brought closer to that of inviscid MHD. This is also evident from the fact
that the $\beta=400$ ``Limiters v1'' run more closely resembles inviscid MHD than its $\beta=100$ equivalent, as it
features a larger fraction of the domain that is susceptible to micro-instabilities.

Interestingly, the suppression of $S$ due to magneto-immutability is somewhat unexpected for the present simulations.
According to \citet{Majeski2024}, magneto-immutability is predicted to occur only in sub-Alfv\'enic turbulence, which
represents a small fraction, if any, of the scales resolved by this study. To be certain of magneto-immutability's
presence, it must be considered whether other signatures of this self-organization effect can be identified as well. One
such signature, and perhaps magneto-immutability's most important consequence, is that it limits the
thermalization of turbulent kinetic energy through viscous dissipation, which itself is proportional to $S$. In
\citet{squire2019}, \citet{Squire2023}, and \citet{Majeski2024}, this suppression of viscous dissipation was found to be
so strong in weakly collisional and collisionless turbulence simulations that it rendered the cascades nearly
conservative. Therefore, to understand just how magneto-immutable the present simulations are, we must further evaluate
the amount of heating that results from the Braginskii viscous stress.

\subsection{Transfer Functions}\label{sec:transfer_functions}

To measure viscous dissipation, we compute energy-transfer functions in the manner of \citet{Grete2017}, \citet{Arzamasskiy2023},
\citet{Squire2023}, and \citet{Majeski2024}, focusing on the transfer of the kinetic
energy of the gas flows into the internal energy $\rm{U} = \rho\epsilon$, via the Braginskii stress. The transfer function is defined as
\citep[][]{Squire2023,Majeski2024}: 
\begin{equation}
{\cal T}_{q{\rightarrow}k}^{\Delta{p}\rm{U}} = \int{{\rm{d}^3{\bf x}}}\langle\sqrt{\rho}{\bf v}\rangle_k\cdot\frac{\bf B}{\sqrt{4\pi\rho}}\cdot\nabla\left\langle\frac{\Delta{p}}{B^2}{\bf B}\right\rangle_q.
\end{equation}
This is a 2D transfer function from the wavevector shell $q$ to another shell $k$; to determine the net total transfer into
(${\cal T}^{\Delta{p}\rm{U}} > 0$) or out of (${\cal T}^{\Delta{p}\rm{U}} < 0$) the shell $k$ from or to all other scales, we sum over all $q$, obtaining the 1D transfer function ${\rm T}^{\Delta{p}\rm{U}}(k)$ \citep[][their Equation
3.10]{Squire2023}:

\begin{equation}
{\rm T}^{\Delta{p}\rm{U}}(k) = {\sum_q}~{\cal T}_{q{\rightarrow}k}^{\Delta{p}\rm{U}}
\end{equation}

In Figure \ref{fig:transfer_function}, we show this integrated transfer function for all our simulations, evaluated
within a cubical volume of (320~kpc)$^{3}$ centered on the cluster potential minimum. Similarly to Sections
\ref{sec:turb_spectra} and \ref{sec:proj_vel}, we apply a Tukey filter to the data before computing the Fourier
transforms. The curves shown are time-averaged transfer functions ${\rm T}^{\Delta{p}\rm{U}}(k)$ computed for all of the
simulation snapshots during $t \in [3.5, 4.0]$~Gyr. As in \citet{Squire2023} and \citet{Majeski2024}, we normalize ${\rm
T}^{\Delta{p}\rm{U}}(k)$ by the total energy-cascade rate ${\rm T}_{\rm total}(k) \sim E(k)/\tau(k)$, where $E(k)$ is
the kinetic energy density at wavenumber $k$ and $\tau(k) \sim 1/kV_k$ is the eddy turnover time at that scale. With the
transfer functions normalized in this way, a non-zero flat curve represents a contribution that is a constant fraction
of what would be a conservative cascade. For both the $\beta_{\rm ini} = 100$ and $\beta_{\rm ini} = 400$ cases, the
normalized transfer function ${\rm T}^{\Delta{p}\rm{U}}$ depends weakly on $k$ for 0.007~kpc$^{-1} \lesssim k \lesssim
0.1$~kpc$^{-1}$, indicating that the pressure anisotropy contributes a roughly constant fraction of the
energy-cascade rate to the internal energy in the inertial range of the turbulence. Outside this range, the transfer
function steepens at both low and high $k$---for the latter, the numerical dissipation begins to dominate, whereas for the former, the corresponding length scales approach the size of the extraction region.

Comparing the simulations with different limiters, we find that strong dissipation occurs in the ``Unlimited''
simulations for both values of $\beta_{\rm ini}$. In the simulations with hard-wall limiters, the effect on
the dissipation depends on the magnetic-field strength. Consistent with our previous results, for
$\beta_{\rm ini}$ = 100, the magnetic field is strong enough for the distribution of pressure anisotropy to be very similar
regardless of the limiters that are applied, and thus the transfer functions have similar values in the inertial range.
In contrast, in the $\beta_{\rm ini}$ = 400 simulation, the limits in the ``Limiters, v1'' simulation have a significant
effect on the viscous dissipation, but less so in the ``Limiters, v2'' case, which has very similar dissipation to the
``Unlimited'' simulation.

The dependence of the viscous heating on the limiter scheme may be responsible for the different degrees of
steepening observed in the spectra of Figure \ref{fig:vas}. The steepest spectra at each $\beta_{\rm ini}$ correspond to
the simulations for which viscous dissipation was the strongest throughout the inertial range in Figure
\ref{fig:transfer_function}. Furthermore, the difference between the ``Limiters, v1'' and ``Limiters, v2'' spectra is
slightly larger at $\beta_{\rm ini}$ = 400 than at $\beta_{\rm ini}$ = 100, a feature that also appears in the
respective transfer functions. If these viscous stresses are indeed responsible for the observed steepening of the
spectra, then it would in turn imply that a magneto-immutable state has not been fully realized by the turbulence
present in these simulations. Rather, it may be that some form of ``partial magneto-immutability'' is observed
here, which warrants further, dedicated investigation.

\section{Summary and Conclusions}\label{sec:conclusions}

In this work, we have presented simulations of sloshing and turbulent motions in the ICM of a cool-core galaxy
cluster, including Braginskii viscosity with limiters on the pressure anisotropy due to plasma instabilities. Our main
results can be summarized as follows:

\begin{itemize}

\item As shown in previous works, Braginskii viscosity can suppress the growth of KHI at CF surfaces in a sloshing cool
core, though the anisotropy of the associated momentum transport reduces its effect relative to an isotropic viscosity.
If the magnetic field is initially strong, or if hard-wall limiters are applied to the pressure anisotropy, the visible
effect of viscosity can be not much more significant than the effect that magnetic tension already has on KHI at CFs.
The differences between simulations with or without the hard-wall limiters are less significant in the case where
$\beta_{\rm ini} = 100$ because stronger magnetic fields make plasmas more resilient against microscale instabilities.

\item The sloshing and turbulent motions in the core region produce pressure anisotropies, caused by rapid changes in
the magnetic field's strength, with the maximum absolute values of the pressure anisotropy $|\delta_p| \sim 0.03-0.07$
at $\beta \sim 100-1000$ (Section \ref{sec:anisotropy_results} and Figure \ref{fig:beta_phase}), corresponding to $\beta|\delta_p| \sim 3-70$, indicating that there are regions where the effect of the pressure anisotropy can become
very large compared to the magnetic tension. In the simulations where hard-wall limiters are not applied, the most
significant regions of positive pressure anisotropy are located along the CF surfaces where the magnetic field
amplification is strongest, whereas negative pressure anisotropies are most prominent in regions of adiabatic expansion.
In some of these regions, the relative importance of the pressure anisotropy over the magnetic tension can be very
large, with the ratio $|\Theta| \sim 10^2-10^3$ (see Equation \ref{eqn:theta}). The simulations have a slight preference
for positive pressure anisotropy, with $\sim$52-65\% of the volume having $\Delta{p} > 0$. 

\item In the simulations without hard-wall limiters to model the effect of plasma instabilities, anywhere from 1\%
to 30\% of the volume can fall outside of one of these instabilities' thresholds, depending on the instability in question and the magnetic field's
strength. When the firehose and/or mirror limiters are applied, the viscous stress arising from the pressure anisotropy
throughout the volume is comparable to, or smaller than, the magnetic tension, by construction. A small but
non-negligible portion of the volume has values of the pressure anisotropy right at the mirror and (oblique-)firehose
thresholds, which in the former case may have dramatic implications for cosmic-ray transport in the ICM, due to
scattering by micro-mirrors \citep{Reichherzer2025}. These regions are small and scattered outside of and in between the
regions where sloshing has significantly amplified the magnetic field. In the simulations where the less-restrictive
ion-cyclotron instability threshold is used to limit the Braginskii stress, the range of positive pressure anisotropies that is allowed is very similar to the case without any limiter at all. 

\item The turbulence driven by the sloshing motions tangles the magnetic field, which reduces the effective viscosity in
the core region due to the anisotropy of the viscous momentum flux with respect to the magnetic field's local direction.
As to the reduction in viscosity resulting from the application of the hard-wall limiters, this effect is most prominent
if the magnetic field is weaker, so it occurs mostly outside the sloshing regions, in which shear motions and turbulence
amplify the magnetic field and make the pressure-anisotropic plasma more stable. Outside these regions, the viscous flux
can be strongly suppressed, especially in our simulations where the mirror threshold provides the limit on positive
values of the pressure anisotropy. In the simulations where the ion-cyclotron threshold is used instead, the viscous
flux is close to the Braginskii value throughout the core. With all of this taken into account, the effective viscosity
in the core region is reduced to $\sim$20\% of the Spitzer value within $\sim$57-75\% of the volume, and to $\sim$50\%
of the Spitzer value within $\sim$90-96\% of the volume.

\item In the absence of any viscosity, the merger-driven sloshing motions at late times produce a turbulent velocity
field with a Kolmogorov-like velocity-amplitude spectrum in 3D, with $A_{\rm 3D}(k) \propto k^{-1/3}$. Adding Braginskii
viscosity steepens the slope of the spectrum to $A_{\rm 3D}(k) \propto k^{-2/3}$. The most natural interpretation of
this result is that Braginskii viscosity damps turbulent motions, but relatively ineffectively, as its ability to do so
is moderated through a combination of micro-instability scattering and self-organized magneto-immutability of the
turbulence. When limiters on the pressure anisotropy are applied, the spectral slope remains as steep as without them,
though the characteristic magnitude of the velocities is not as strongly suppressed as in the case without the limiters.

\item Qualitatively, the projected line width and shift across all the simulations look very similar on large scales,
but small-scale features are suppressed in both quantities compared to the inviscid runs, consistent with the steeper
velocity-amplitude spectra. The 2D spectra of the line shift also show a difference in slope between
inviscid and viscous simulations, but the difference in the slopes is more difficult to discern due to projection
effects.

\item We also computed the amplitude spectra of density fluctuations in the same manner as has been done in
observational studies, from SB fluctuations in an X-ray image. In this case, the differences between the simulations
with different prescriptions for viscosity are very difficult to discern, due to the smoothing of the amplitude spectra
by the $\Delta$-variance filter.

\item By comparing PDFs of the parallel rate of strain across all runs, we found that these distributions in Braginskii
simulations were always significantly narrower than those from the corresponding inviscid MHD simulations. This is a key
characteristic of the self-organization process known as magneto-immutability, which enhances the effective viscosity
suppression beyond that which results from tangled field lines alone. These signatures were found to be weakest in
simulations where the hard-wall micro-instability limiters were most restrictive, likely due to the fact that stronger
microinstability scattering drives the Braginskii model toward inviscid MHD.

\item We also computed the energy transfer functions from the turbulent flows to the internal energy (due to dissipation
by Braginskii stresses), to quantify the effect of the pressure anisotropy on the energy cascade rate. The transfer
functions derived from the simulations are a roughly constant fraction of the total energy cascade rate in the inertial
range. The differences between the simulations with different limiters are most pronounced in the cases where the
magnetic field is weaker and the instability-based limiters are more restrictive. These transfer functions may be
responsible for the observed steepening of the turbulent kinetic energy spectra, and suggest that the form of
magneto-immutability observed in these simulations may only be a partial realization of the self-organizing effect.

\end{itemize}

In general, we show that if viscous momentum transport in the ICM is controlled by the pressure anisotropy with respect
to the field lines, its effects on observed properties in the ICM can be very subtle. Given its anisotropy due to the
magnetic field and limits imposed on it from plasma instabilities, the viscous stress will be comparable to the magnetic
tension. For this reason, it is difficult to constrain the effective viscosity in the ICM from observations of KHI at CF
surfaces (since these are also affected by magnetic tension), except perhaps to rule out extremely low effective
Reynolds numbers \cite[see, e.g., the dramatic effect of full isotropic Spitzer viscosity on the smoothness of CF
surfaces in][]{ZuHone2010,ZuHone2015,Hsieh2026}. 

The effects of viscosity may also be detected in the properties of the ICM velocity field, which is newly accessible
with microcalorimeter instruments like {\it XRISM}'s Resolve. In our simulations, the 3D velocity-amplitude spectra show
a steeper slope than a Kolmogorov spectrum when Braginskii viscosity is included (this is also observed in our 2D
projected maps of the emission-weighted mean velocity, but with less clarity). One of the unresolved questions from our
work is the physical origin of this steep slope. In Section \ref{sec:turb_spectra}, we suggested that it may be simply
reflective of the underlying sloshing motions without significant modification by the turbulence, or it may have a
similar character to the steep slopes seen in simulations of the turbulent MHD dynamo. It must be said that it is not
completely straightforward to compare our results to those from the simulations where turbulence is driven in
non-stratified boxes. Aside from the effects of stratification in our simulations, there is the fact that our
simulations are not continuously stirred, but the sloshing motions are merger-driven. These motions decay into
turbulence at different rates depending on how far from the cluster center they are located, and turbulence at smaller radii is dissipated
earlier on in the simulation, as new turbulence is generated from the expanding cold front motions at larger radii. In order to pin down more conclusively the physics behind the slope of these velocity-amplitude spectra with different viscosities, it will be necessary to perform new numerical simulations of driven MHD turbulence in stratified boxes, with higher spatial resolution than ours, in order to reduce numerical dissipation and extend the inertial range of the turbulence. 

This work employed some simplifications in the physical modeling of ICM that could be improved upon in future work. For
example, Equation \ref{eqn:pressure_anisotropy} assumes that the pressure anisotropy can be obtained at all times by
balancing its production by adiabatic invariance with isotropization by collisions. A more accurate approach would
self-consistently evolve the pressure anisotropy with time without assuming this equality, using the CGL-MHD equations
\citep{Squire2023,Majeski2024}. Using this approach, we may see differences from the simulations presented here in
regions where the magnetic field is increasing or decreasing rapidly compared to the ion-ion collision frequency, though
this may not be an important consideration in cluster cool cores. We may also add the effects of thermal conduction,
which \citet{Squire2023} showed can smooth out the pressure anisotropy along the magnetic field lines and thus also
determine how much of the plasma has pressure anisotropy at the instability thresholds and beyond. 

Finally, while this manuscript was in preparation, \citet{Hsieh2026} presented similar simulations of sloshing cool cores with different models for viscosity, including Braginskii viscosity and hard-wall limiters, and different initial magnetic-field strengths. In contrast to our work, they also included models with isotropic viscosity, and their finest cell size is somewhat coarser, with $\Delta{x} = 3.9$~kpc. Though the analyses carried out in our work and theirs are not identical, where they are similar, they are consistent with each other. For example, they find that while Braginskii viscosity has a suppressing effect on the appearance of KHI at CF surfaces, it is modest (especially when compared with simulations with isotropic viscosity), and that the degree of this suppression depends on whether the hard-wall limiters are implemented and the initial magnetic-field strength implemented in the simulation.

\begin{acknowledgments}
JAZ thanks Thomas Berlok, Tirso Marin-Gilabert, and Franco Vazza for useful discussions. Calculations were performed
using the computational resources of the Advanced Supercomputing Division at NASA/Ames Research Center, and the Harvard
University Faculty of Arts and Sciences Research Computing. Support for JAZ was provided by the {\it Chandra} X-ray
Observatory Center, which is operated by the Smithsonian Astrophysical Observatory for and on behalf of NASA under
contract NAS8-03060. Support for SM was provided by NSF CAREER Award No.~1944972, the Charlotte Elizabeth Proctor
Fellowship at Princeton University, and NASA Astrophysics Program grant 80NSSC22K0828. The work of AAS was supported in
part by the UK STFC (grant ST/W000903/1) and by the Simons Foundation via a Simons Investigator Award. AH and IZ were 
partially supported by NASA award 80NSSC24K1488, and NASA/Chandra awards GO1-22123A and AR4-25012X.
\end{acknowledgments}

\begin{contribution}
The running and analysis of the simulations presented in this work, as well as the majority of the writing, were carried
out by JAZ. SM provided the calculations and software for the transfer functions, suggestions for interpretation and
analysis, contributed to the writing, and reviewed the manuscript. AH provided the calculations of the SB residual maps
and the amplitude spectra of their fluctuations. FL, AAS, and IZ provided suggestions for interpretation and analysis,
contributed to the writing, and reviewed the manuscript.
\end{contribution}

\software{astropy \citep{Astropy2013,Astropy2018,Astropy2022},  
yt \citep{Turk2011}, matplotlib \citep{Hunter2007}, numpy \citep{Harris2020}, scipy \citep{Virtanen2020}.}
     
\appendix

\section{Initial Relaxation of the Magnetic Field}\label{sec:magnetic_relaxation}

\begin{figure*}
\centering
\includegraphics[width=0.98\textwidth]{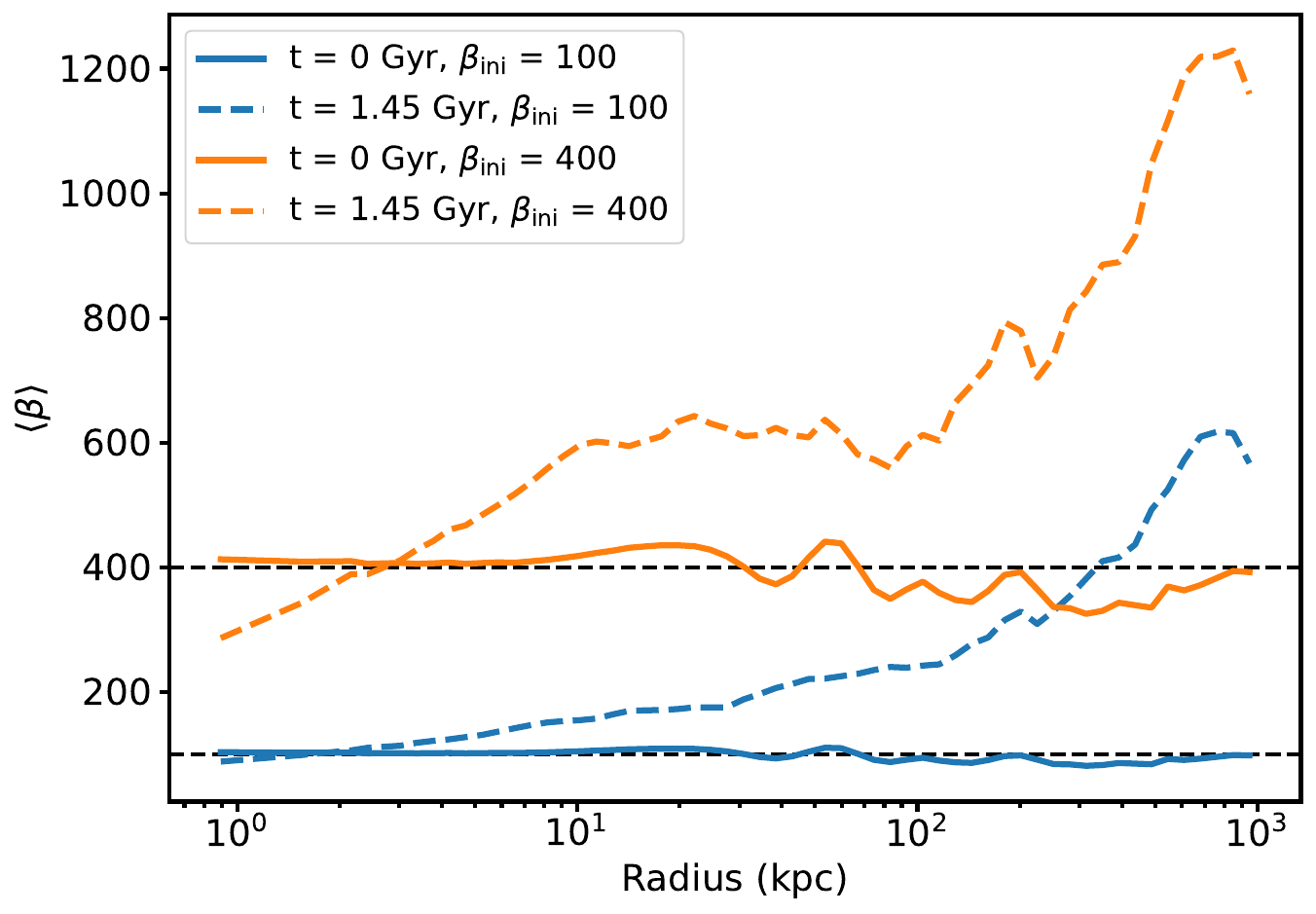}
\caption{Azimuthally averaged profiles of the plasma $\beta$ at $t$ = 0 and $t$ = 1.45~Gyr for the two initial values of $\beta_{\rm ini}$.\label{fig:beta_profile}}
\end{figure*}

As noted in Section \ref{sec:ICs} and \citet{ZuHone2011a}, the initial magnetic field is not an equilibrium field (it is
not force-free), and so during the approach of the subcluster, the magnetic field relaxes to a lower-energy state. In
this process, energy is transferred from the magnetic field to gas motions. The energy of these motions is not
significant, but the more important consideration for our purposes is the fact that the magnetic pressure decreases,
which is especially relevant for the viscous simulations with hard-wall limiters that are based on it.

To quantify the importance of this effect, in Figure \ref{fig:beta_profile} we show azimuthally-averaged radial profiles
of the plasma $\beta$ at $t = 0$ and $t = 1.45$~Gyr (shortly before the pericenter passage) for both values of
$\beta_{\rm ini}$. For both simulations, the plasma $\beta$ indeed increases between these times, to $\beta \sim$ 200
for the $\beta_{\rm ini}$ = 100 simulation and to $\beta \sim 600$ for the $\beta_{\rm ini}$ = 400 simulation, within a
radius of $\sim$100~kpc. At larger radii, the plasma $\beta$ is even higher, but caution should be taken in the
interpretation as these are the radii where the subcluster has already heated and compressed the gas along one direction
(included in the average), as well as increased the magnetic pressure through flux freezing.

We see from this analysis that in the region where the sloshing motions are most prominent in both simulations, the
decrease in the magnetic pressure is the lowest, and the two simulations still have a significant difference in magnetic-field strength. Thus, this initial magnetic-field evolution does not significantly affect our results or the value
of the comparison between the two initial conditions for the magnetic field.

\bibliography{ms}{}
\bibliographystyle{aasjournalv7}

\end{document}